\documentclass{birkjour}
\usepackage{amsmath}
\usepackage{amsfonts}
\usepackage{amssymb}
\usepackage{graphicx}

\usepackage{amsthm}
\usepackage{verbatim}

\newtheorem{defn}{Definition}
\newtheorem{thm}{Theorem}
\newtheorem{prop}{Proposition}
\newtheorem{lemma}{Lemma}

\theoremstyle{definition}
\newtheorem{rmk}{Remark}

\begin{document}

\title[Localized stable manifolds]{Localized stable manifolds for whiskered tori in coupled map lattices with decaying interaction}
\author[D. Blazevski]{Daniel Blazevski}
\address{Institute for Mechanical Systems\\
Department of Mechanical and Process Engineering\\
ETH Zurich\\
Tannenstrasse 3\\
8092 Zurich\\
Switzerland
}
\email{blazevski@imes.mavt.ethz.ch}
\author[R. de la Llave]{ Rafael de la Llave}
\address{School of Mathematics\\
Georgia Institute of Technology\\
686 Cherry St.\\
Atlanta GA 30332 \\
USA 
}
\email{rafael.delallave@math.gatech.edu}

\begin{abstract}
In this paper we consider lattice systems coupled by local interactions.  We prove invariant manifold theorems for
whiskered tori (we recall that whiskered tori are quasi-periodic solutions with exponentially contracting and expanding
directions in the linearized system).  The invariant manifolds we construct generalize the usual (strong)
(un) stable manifolds and allow us to consider also non-resonant manifolds.  We show that if the whiskered tori
 are localized near a collection of specific sites, then so are the invariant manifolds.

We recall that the existence of localized whiskered tori has recently been proven for symplectic maps and flows
in \cite{FLS}, but our results do not need that the systems are symplectic.  For simplicity we will present first the main
results for maps, but we will show tha the result for maps imply the results for flows.  It is also true that the results
for flows can be proved directly following the same ideas.

\end{abstract}

\maketitle

\section{Introduction}
In this paper we study stable manifold theorems in coupled map lattices.  We recall that coupled map lattices are copies of
a dynamical system at each point in the lattice coupled by some local interaction.  They have been extensively 
studied as models in neuroscience, chemistry and other disciplines
\cite{BraunK98, Jiang_Pesin, Pego1, DauxoisRAW02,  Braun, FloriaBG05,PesinY04,ChazottesF05, BilkiENW07,
GallavottiFPU,Kaneko93}.

The goal of this paper is to prove invariant manifold theorems for localized whiskered tori.  We recall that
whiskered tori are quasi-periodic solutions such that the linearized dynamics around them have exponentially
expanding (or contracting) directions.  In this paper we refer to quasi-periodic as solutions whose frequency 
vector is possibly infinite dimensional
 (sometimes in the literature solutions with an infinite dimensional frequency vector are called almost periodic solutions). On the other side of 
the spectrum, periodic solutions are a particular 
case of quasi-periodic with our definition. 
See Definition \ref{whiskereddef} for a precise definition of a whiskered torus.  

We say that the whiskered tori are ``localized'' when the oscillations are concentrated near a specific
collection of sites (which we allow to be finite or infinite, see Definition \ref{whiskereddef}).  
The existence of localized whiskered tori has been proved (under some hypotheses) for symplectic maps and flows in
\cite{FLS} (e.g. for coupled pendula or Fermi-Pasta-Ulam systems \cite{FPU} ).  Our framework allows us to deal
with general Hamiltonian systems with infinite degrees of freedom with long range (including infinite range) localized 
interactions.

A very paradigmatic example of our framework 
 are Hamiltonian systems, (appearing as discretizations 
of Klein-Gordon equations)  whose energy is given by
\begin{equation} \label{Klein-Gordon} 
H(q, p) = \sum_{n \in \mathbb{Z}^N} \left(\frac{1}{2} p_n^2 + W(q_n) \right) + 
\sum_{j \in \mathbb{Z}^N} \sum_{n \in \mathbb{Z}^N} V_j(q_n - q_{n + j})
\end{equation}
where we have assumed that the system
\begin{equation}
\ddot{q} + W'(q) = 0
\end{equation}
has a hyperbolic fixed point.  This assumption yields whiskered tori for the uncoupled system,
 i.e. the system without the interacting terms $V_j(q_n - q_{n + j})$.  Moreover, suppose that the coupling 
potentials $V_k$ satisfy
\begin{equation}
\| V_k \|_{C^2_\rho} \leq C_V \Gamma(k)
\end{equation}
where $\Gamma: \mathbb{Z}^N \rightarrow \mathbb{R}$ is a decay function, a notion quantifiying fast decay  defined in Section \ref{decay_section}.  
Finite range interactions (e.g. the Frenkel-Kontorova model) of any arbitrary range are included in this example. 
The key novelty is we are able to use the decay properties assumed on the flow or map 
to construct invariant manifolds constructed with decay properties: The manifolds are parameterized by a function $W$ such that
\begin{equation}
 \left| \frac{\partial W_i}{\partial x_j} \right|
\end{equation}
decays as the lattice site $ i \in \mathbb{Z}^N$ gets further away from the excited sites, or
if $i$ is very different from $j$.  See Section \ref{fnspace_embeddings}
for precise definitions of localized embeddings.  

In \cite{FLS} it was shown that 
a whiskered torus with decay exists for the coupled system under the assumption that an approximate whiskered
torus exists. In particular, when the coupling is small, the tori of the uncoupled system persist. 
In this paper, we assume that a whiskered torus exists and show that they have stable and unstable manifolds
that are localized near the torus.    
The methods of this paper allow one to construct whiskers in 
the tori produced in \cite{FLS}. 

Another model very similar in spirit is the coupled 
standard map introduced and explored 
numericaly in \cite{Kaneko85b}. The model was also 
considered in \cite{FLS} and the existence of localized 
quasi-periodic whiskered solutions was established for 
certain values of the regime.  The paper \cite{Kaneko85b} presents 
numerical evidence of analogues of Arnol'd diffusion, which served 
as motivation to understand the stable and unstable manifolds of 
whiskered tori in these systems. A natural step, which we have not 
yet taken is to study the global properties of the manifolds whose
existence is established here. 

We note, however, that in this paper we can consider more general systems since we do not need the 
assumption that the dynamics preserves a symplectic form. 
 Localized quasi-periodic solutions happen also in coupled dissipative
 systems with limit cycles.  Such localized
excitations have been considered in neuroscience (\cite{Iz, Er})
and in other disciplines (\cite{DauxoisPW92,PS04,Peyrard04}). 
Our results also apply in the  dissipative
context and, according to our definition, periodic solutions are 
a particular case of quasi-periodic. 

The main result of this paper is 
an invariant manifold theorem for such localized whiskered tori.  
We show that corresponding to some
spectral subspaces in the linearization, one can find smooth manifolds of initial conditions which converge to the 
quasi-periodic solutions.  The invariant manifold theorem we prove includes as particular cases the classical stable
and unstable manifold theorems or the strong (un) stable manifolds theorems.  We just need some non-resonance conditions
in the spectrum.  This allows us to make sense of the slow manifold in some cases.  
We also prove smooth dependence on parameters, which could 
serve to develop a perturbation theory.

One motivation for the construction of whiskered tori is that they separate asymptotic dynamics.  Transverse 
intersections of stable and unstable manifolds of whiskered tori were constructed for
specific examples in \cite{Arnold64} and conjectured to be a generic mechanism 
 of transport in phase space and global instability.  The mechanism in \cite{Arnold64} has proven 
 to be robust in its goal for finite dimensional systems.
 (see, for example, \cite{DLS} for
a review with references to the original literature in recent developments).  One can hope that similar effects
happen in lattice systems and this paper is a step towards the implementation of the \cite{Arnold64} program in 
coupled map lattices.  Studies of the Arnold mechanism in 
coupled map lattices were undertaken in \cite{Kaneko85b}.

In the applied literature there are quite a number of phenomena (e.g.
 ``bursting'' \cite{Bursting}, ``spiking'' \cite{spiking},
``transfer of energy'' \cite{MezicPG, MezicCO}) which indeed are reminiscent of homoclinic chaos in infinite dimensions. 
We think it would be interesting to clarify mathematically these issues.
The localization properties of invariant manifolds has relevance for 
the study of statistical mechanics. 

Stable and unstable manifolds, of general sets also play a role in 
the study of spatio-temporal chaos \cite{BunimovichS88}.
 In this paper we consider only 
manifolds of quasi-periodic orbits since the manifolds are more 
differentiable. Some different results for more general sets 
can be found in \cite{FLM2}.

This paper is organized as follows:  Section \ref{prelim_section} consists of technical definitions for 
the setup of our results.  
In particular, we define the phase space we use, localized interactions, and analytic embeddings of
localized whiskered tori and stable and unstable manifolds.

In Section \ref{statement_section} we provide statements of our results.  We start by stating Theorem \ref{maintheorem}, which
assumes a more classical notion of a whiskered torus.  Then we state 
 Theorem \ref{maintheorem_spec}, which, given our methods, is a natural generalization of Theorem \ref{maintheorem}.
Theorems \ref{maintheorem} and \ref{maintheorem_spec} are results for discrete maps on lattices.  In section 
\ref{flowsection} we show that Theorem \ref{maintheorem}
implies an analogous result for flows on lattices, which is the content of Theorem \ref{maintheorem_flows}.  Section 
\ref{proof} contains a proof of Theorem \ref{maintheorem_spec}.    

\section{Preliminaries: the Phase Space and Functions with Decay}\label{prelim_section}
In this section we introduce several technical definitions that follow the setup in \cite{FLS, FLM1, FLM2}.  This section
can be used as a reference.  The central idea is to make precise the notion that objects are localized by imposing that 
the derivatives of a component with respect to a variable are small if the distance between index of the component and 
the variable is large.   To avoid unnecessary repetitions, but to 
maintain some readability, we note that the definitions of 
Sections~\ref{phase_space_section},\ref{decay_section}, \ref{dyn_space}
are the same as in \cite{FLS, FLM1, FLM2} (even if we suppress the 
references to symplectic forms,etc. in \cite{FLS}). 
The subsequent sections are new. In particular, we need some extra 
definitions to deal with infinite dimensional tori and the embeddings of the invariant manifolds associated to the tori as such objects were
not considered in \cite{FLS, FLM1, FLM2}.  

We need two sets of definitions of localized objects:  diffeomorphisms and the embeddings giving the parameterization
of the invariant manifolds.  An important technical notion introduced in \cite{SRB} is that of a ``decay function''.

With the technical definitions defined in Sections \ref{phase_space_section} - \ref{spec_form_defn} , 
we will see that some (but not all) of the techniques from finite dimensional systems 
generalize to the infinite dimensional setting of coupled map lattices.  Of course, some features have to be 
significantly different.  For example, coupled map lattices may have uncountably many periodic points which are
uniformly hyperbolic, as well as other features that are impossible in finite dimensional differentiable systems 
in a compact manifold.  
Hence, one has to give up either differentiability or local compactness of phase space. 
Several compromises are possible and, it is possible to give topologies 
that keep compactness but give up  differentiability, which is convenient for 
ergodic arguments (see e.g. \cite{Jiang_Pesin,Rugh02} ).
Since in this paper we  will be performing a geometric analysis,
the set-up of this paper emphasizes differentiability.

To keep the differentiability assumption, it
is natural to model phase space in $\ell^\infty$, but since $\ell^\infty$ is not reflexive, 
this opens some new  difficulties, which we will have to overcome.

In what follows, we use the notation $\ell^\infty(\mathbb{Z}^N; X)$ to denote the space of bounded sequences of elements in a
Banach space $X$ indexed by $\mathbb{Z}^N$
\begin{equation}
 \ell^\infty(\mathbb{Z}^N; X) = \left\{  ( x_i ) | x_i \in X, \sup_{i \in \mathbb{Z}^N} \| x_i \| < \infty \right\}
\end{equation}

\subsection{The Phase Space}\label{phase_space_section}
In this section we will define the phase space of the system we will be considering.   

The phase space for each lattice site will be 
$M = \mathbb{T}^l \times \mathbb{R}^{d}$, where $\mathbb{T} = \mathbb{R}/\mathbb{Z}$.  This choice of $M$ is 
done for convenience since $\mathbb{T}$ has straight-forward complex extensions and, since we are considering neighborhoods of
 quasi-periodic solutions, it entails no loss of generality.  The full
phase space $\mathcal{M}$ for the entire lattice system is a subset of
\begin{equation}
 M^{\mathbb{Z}^N} = \prod_{j \in \mathbb{Z}^N} M
\end{equation}
consisting of bounded sequences of points in $M$.  That is,
\begin{equation}
 \mathcal{M} = \ell^\infty(\mathbb{Z}^N; M) = \left\{ x \in M^{\mathbb{Z}^N} : \sup_{i \in \mathbb{Z}^N} | x_i | < \infty \right\}
\end{equation}
unless $l = 0$ $\mathcal{M}$ will not be a Banach space, but will be a Banach manifold. 

Unless otherwise specified we will write $\ell^\infty(\mathbb{Z}^N)$ to mean $\ell^\infty(\mathbb{Z}^N; M)$.
The space $\mathcal{M}$
has a natural notion of distance, which is given by
\begin{equation}
 d(x, y) = \sup_{i \in \mathbb{Z}^N} d(x_i, y_i)
\end{equation}
Although $\mathcal{M}$ is a manifold, since $\mathbb{T}^l$ is a Euclidean space (i.e. we can identify the tangent space at each point with 
$\mathbb{R}^l$) the tangent space of $\mathcal{M}$ can be identified with $\ell^\infty$.  

Since we want to consider analytic functions defined on $\mathcal{M}$ it is natural to consider the complexification of $\mathcal{M}$, which is given
by
\begin{equation}
 \mathcal{M}^{\mathbb{C}} = \left\{ z \in \prod_{j \in \mathbb{Z}^N} M^{\mathbb{C}} : \sup_{i \in \mathbb{Z}^N} |z_i| < \infty \right\}
\end{equation}
Unless otherwise specified, we will be working with $\mathcal{M}^{\mathbb{C}}$ and omit the superscript $\mathbb{C}$.  
We use the $\ell^\infty$ norm to allow for components of the tori to be uniform in size irrespective of
the lattice site.  Using, for example, the $\ell^2$
 norm would require that the components of the tori vanish at infinity.  The $\ell^\infty$ norm is also
convenient for the notion of decaying interaction we use in the following sections.  In this paper, we will consider only analytic function and not 
$C^r$ functions.  

\subsection{Decay Functions and Corresponding Function Spaces}\label{decay_section}

We will now discuss suitable notions of decaying interactions, and the appropriate function spaces that are used throughout 
the paper.  As mentioned before,
we will assume that the coupling of the lattice sites is localized.  
To make this notion more precise, we will use the notion of a decay function as done in \cite{SRB, FLM1, FLM2, FLS}.
\begin{defn}
A function $\Gamma: \mathbb{Z}^N \rightarrow \mathbb{R}_+$ is a decay function provided that
\begin{equation} \begin{split}
&\sum_{j \in \mathbb{Z}^N} \Gamma(j) \leq 1 \\
&\sum_{j \in \mathbb{Z}^N} \Gamma(i - j)\Gamma(j - k) \leq \Gamma(i - k) \text{   }, i, k \in \mathbb{Z}^N
\end{split}\end{equation}
\end{defn}

Given a decay function $\Gamma$ we consider several spaces of functions that ``decay like $\Gamma$''.  We will have 
two types of functions:  the ones that give the dynamics and functions for the parameterization of invariant objects.    
Roughly, the idea of spaces of functions with decay is that the influence of site $i$ on site $j$ is bounded by 
$C \Gamma(i - j)$.  The influence is measured as the size of the partial derivative of the $i$-th component with respect to
the $j$-th variable.  

\subsubsection{Function Spaces for the Dynamics}\label{dyn_space}
In this section we will discuss the function spaces relevant for the map $F$ that governs the dynamics of the
lattice.  

%This is similar to the setup in \cite{FLS}, though we will need to consider analytic functions of infinitely
%many variables and some notions that we consider were either only mentioned briefly or not at all in \cite{FLS}.  

First, we consider the Banach space of decay linear operators that are represented by their matrix elements

\begin{equation}\begin{split}
& \mathcal{L}_\Gamma(\ell^\infty(\mathbb{Z}^N)) = \\
& \begin{Bmatrix} 
& A \in \mathcal{L}(\ell^\infty(\mathbb{Z}^N)) : 
\text{ for every } i, j \in \mathbb{Z}^N \exists A_{i j} \in \mathcal{L}(M), \\
& (Au)_i = \sum_{j \in \mathbb{Z}^N} A_{i j} u_j,  \sup_{i, j \in \mathbb{Z}^N} \Gamma(i - j)^{-1} |A_{i j}| < \infty \end{Bmatrix}
\end{split}\end{equation}
where $\mathcal{L}(\ell^\infty(\mathbb{Z}^N))$ denotes the usual space of continuous linear maps
from $\ell^\infty(\mathbb{Z}^N)$ to itself.  A norm on $\mathcal{L}_\Gamma(\ell^\infty(\mathbb{Z}^N))$ is given by
$$\|A\|_\Gamma = \sup_{i, j \in \mathbb{Z}^N} \Gamma(i - j)^{-1} |A_{i j}|$$

\begin{rmk}  As emphasized in \cite{FLM1},
not all bounded linear operators from $\ell^\infty(\mathbb{Z}^N)$ can be represented by their matrix elements.  For example consider the linear closed subspace
$E_0 = \{ v \in \ell^\infty(\mathbb{Z}) | \lim_{|j| \rightarrow \infty} v_j \text{ exists } \}$ of
$\ell^\infty(\mathbb{Z}, \mathbb{R})$ and the bounded linear functional
$f: E_0 \rightarrow \mathbb{Z}$ defined by
\begin{equation}
f(v) = \lim_{|j| \rightarrow \infty} v_j 
\end{equation}
The linear operator $f$ is bounded, having operator norm equal to $1$. By the Hahn-Banach theorem we can extend $f$ to a bounded linear functional $L$ on all of $\ell^\infty(\mathbb{Z})$ which also has norm $1$.  
The matrix elements of $L$ are zero, yet certainly $L$ is a non-zero functional.  
\end{rmk}

When we consider functions in complex domains, the derivatives 
are understood to be complex derivatives. When we consider 
functions from a Banach space, the derivatives are understood 
to be the strong derivatives. 

The space of $C^1$ functions on a open set $\mathcal{B} \subset \mathcal{M}$ that decay like $\Gamma$ is
\begin{equation}
C^1_\Gamma(\mathcal{B}) = 
\begin{Bmatrix} 
& F: \mathcal{B} \rightarrow \mathcal{M} : F \in C^1(\mathcal{B}), DF(x) \in C^0( 
\mathcal{B}, \mathcal{L}_\Gamma(\ell^\infty(\mathbb{Z}^N)) )  \\
& \sup_{x \in \mathcal{B}} \| F(x) \| < \infty, \sup_{x \in \mathcal{B}} \| DF(x) \|_\Gamma < \infty
\end{Bmatrix}
\end{equation}
The space $C^1_\Gamma(\mathcal{B})$ is a Banach space with the norm
$$\| F \|_{C^1_\Gamma} = \max \left( \sup_{x \in \mathcal{B}} \| F(x) \|, \sup_{x \in \mathcal{B}} \| DF(x) \|_\Gamma \right) $$

\begin{defn}
Let $\mathcal{B}$ be an open set of $\mathcal{M}$.  We say that $F: \mathcal{B} \rightarrow \mathcal{M}$ is analytic and decasys like $\Gamma$
if it is in $C^1_\Gamma(U_r)$, where $U_r$ is a complex neighborhood of $\mathcal{B}$.
\end{defn}

We will also need to consider the space $\mathcal{L}^k(\ell^\infty(\mathbb{Z}^N))$ of $k$-multilinear maps that are represented
by their matrix elements, that is $B \in \mathcal{L}^k(\ell^\infty(\mathbb{Z}^N))$ if and only if we can write
\begin{equation}
 (B(x^1, \ldots, x^k))_i = \sum_{(i_1, \ldots, i_k) \in (\mathbb{Z}^N)^k} B_{i, i_1, \ldots, i_k}x^1_{i_1} \cdots x^k_{i_k}
\end{equation}
where $i, i_1, \ldots, i_k \in \mathbb{Z}^N, (x^1, \ldots, x^k) \in (\ell^\infty(\mathbb{Z}^N))^k$ and $B_{i, i_1, \ldots, i_k} \in \mathcal{L}^k(M, M)$.
Given a decay function $\Gamma$, we will consider the space $\mathcal{L}^k_\Gamma(\ell^\infty(\mathbb{Z}^N))$ of $k$-multilinear
maps given by their matrix elements that decay like $\Gamma$, that is all maps $B \in \mathcal{L}^k(\ell^\infty(\mathbb{Z}^N))$ such that
\begin{equation}
 |B_{i, i_1, \ldots, i_k} | \leq C \min (\Gamma(i - i_1), \ldots, \Gamma(i - i_k))
\end{equation}
for some $C > 0$.  A norm on $\mathcal{L}^k_\Gamma(\ell^\infty(\mathbb{Z}^N))$ is given by
\begin{equation}
 \| B \|_\Gamma = \sup_{i, i_1, \ldots, i_k \in \mathbb{Z}^N} |B_{i, i_1, \ldots, i_k} | \max (\Gamma^{-1}(i - i_1), \ldots, \Gamma^{-1}(i - i_k)).
\end{equation}

\begin{lemma}\label{complinear}
(1) If $A, B \in \mathcal{L}_\Gamma(\ell^\infty(\mathbb{Z}^N))$ then $AB \in \mathcal{L}_\Gamma(\ell^\infty(\mathbb{Z}^N))$ and
$$\| AB \|_\Gamma \leq \|A\|_\Gamma \|B \|_\Gamma$$
(2) More generally, if $A \in \mathcal{L}^k_\Gamma(\ell^\infty(\mathbb{Z}^N))$ and 
$B_j \in \mathcal{L}^{n_j}_\Gamma(\ell^\infty(\mathbb{Z}^N))$
for $1\leq j \leq k$.  Then the contraction 
$AB_1 \cdots B_k \in \mathcal{L}_\Gamma^{n_1 + \cdots + n_k}(\ell^\infty(\mathbb{Z}^N))$ defined by
$AB_1 \cdots B_k(v_1, \cdots, v_k) = A(B_1 v_1, \cdots B_k v_k)$ where $v_i \in \ell^\infty(\mathbb{Z}^N)^{n_i}$
satisfies
$$\|AB_1 \cdots B_k \|_\Gamma \leq \|A \|_\Gamma \|B_1\|_\Gamma \cdots \|B_k \|_\Gamma$$
%(3)if $A \in S^k_{\underline{\rho},\Gamma}(\ell^\infty(\mathbb{Z}^N))$ and $B_j \in S^{n_j}_{\underline{\rho},\Gamma}(\ell^\infty(\mathbb{Z}^N))$
%for $1\leq j \leq k$.  Then the composition $AB_1 \cdots B_k \in S_{\underline{\rho}, \Gamma}^{n_1 + \cdots + n_k}(\ell^\infty(\mathbb{Z}^N))$ and
%$$\|AB_1 \cdots B_k \|_{\underline{\rho},\Gamma} \leq \|A \|_{\underline{\rho}, \Gamma} \|B_1\|_{\underline{\rho}, \Gamma} \cdots \|B_k \|_{\underline{\rho}, \Gamma}$$
\end{lemma}
 This has already been proven in \cite{FLM1}.  

%As for part $(3)$, the fact that $AB_1 \cdots B_k(\theta)$ is
%in $\mathcal{L}^k_\Gamma$ for each $\theta$ is a consequence of $(2)$, and analyticity in $\theta$ follows from the chain and product rule.
%Finally the estimate $\|AB_1 \cdots B_k \|_{\underline{\rho},\Gamma} \leq \|A \|_{\underline{\rho}, \Gamma} \|B_1\|_{\underline{\rho}, \Gamma} \cdots \|B_k \|_{\underline{\rho}, \Gamma}$
%is also a consequence of $(2)$. 
$\Box$

\subsubsection{Function Spaces for Embeddings of Manifolds}\label{fnspace_embeddings}
In this section we will consider spaces of localized vectors and embeddings of invariant manifolds that are used in the paper. 
 We start by discussing the notion of localized vectors and associated multilinear maps.  The tori considered in \cite{FLS}
 are mainly finite dimensional tori and eventually take limits to obtain infinite dimensional tori.  We however, work with 
infinite dimensional tori from the start and therefore state carefully notions of analytic embedding for infinite
dimensional tori and their stable manifolds.  

In general, we will consider a collection $\underline{c} \subset \mathbb{Z}^N$ of ``excited states''.  
We will write $\underline{c} = \{ c_k \}_{k \in \mathcal{K}} \subset \mathbb{Z}^N$,
where $\mathcal{K}$ is a subset of $\mathbb{N}$ used to index the elements
of $\underline{c}$.  The set $\mathcal{K}$ can either be $\mathcal{K} = 1, 2, \ldots, ... n$ for finitely
many excited sites, or  $\mathcal{K} =1, 2, \ldots$ for infinitely many.

\begin{defn}
 Given a decay function and a collection of sites $\underline{c} = \{ c_k \}_{k \in \mathcal{K}} \subset \mathbb{Z}^N$
 for some index set $\mathcal{K}$,
  we define
\begin{equation}
 \| v \|_{\underline{c}, \Gamma} = \sup_{i \in \mathbb{Z}^N} \inf_{k \in \mathcal{K}} | v_i | \Gamma^{-1}(i - c_k)
\end{equation}
we denote
\begin{equation}
 \ell^\infty_{\underline{c}, \Gamma} = \left\{v \in \ell^\infty(\mathbb{Z}^N) | \|v \|_{\underline{c}, \Gamma} \leq \infty \right\}
\end{equation}
That, is, $\ell^\infty_{\underline{c}, \Gamma}$ is the space of vectors localized at the lattice sites $c_k, k \in \mathcal{K}$.

We denote by $\mathcal{L}_{\underline{c}, \Gamma}$ to be the space of linear operators on $\ell^\infty_{\underline{c}, \Gamma}$ such that
\begin{equation}\begin{split}
 &(Av)_i = \sum_{j \in \mathbb{Z}^N} A_{ij}v_j \\
&|A_{ij}| \leq C \min \left( \sup_{k \in \mathcal{K}}\Gamma(i - c_k), \Gamma(i - j) \right)
\end{split}\end{equation}
We denote by $\|A \|_{\underline{c}, \Gamma}$ the best constant C above, i.e.
\begin{equation}
 \|A \|_{\underline{c}, \Gamma} = \max \left( \sup_{i, j \in \mathbb{Z}^N} |A_{ij}|\Gamma^{-1}(i - j), \sup_{i, j \in \mathbb{Z}^N} \inf_{k \in \mathcal{K}}
|A_{ij}|\Gamma^{-1}(i - c_k)\right)
\end{equation}
Finally, we denote by $\mathcal{L}_{\underline{c}, \Gamma}^k$ the space of $k$-multilinear operators $B$ on
 $\ell^\infty_{\underline{c}, \Gamma}$ such that

\begin{equation}
 |B_{i, i_1, \ldots, i_k} | \leq C \min \left(\sup_{k \in \mathcal{K}} \Gamma(i - c_k), \Gamma(i - i_1), \ldots, \Gamma(i - i_k) \right)
\end{equation}
for some $C > 0$.  A norm on $\mathcal{L}^k_{\underline{c}, \Gamma}(\ell^\infty(\mathbb{Z}^N))$ is
given by the best constant $C$ above, that is,
\begin{equation}\begin{split}
 & \| B \|_{\underline{c}, \Gamma}  =  \sup_{i, i_1, \ldots, i_k \in \mathbb{Z}^N} |B_{i, i_1, \ldots, i_k} | \cdot \\ &
 \cdot \max \left( \Gamma^{-1}(i - i_1), \ldots, 
\Gamma^{-1}(i - i_k),  \inf_{k \in \mathcal{K}} \Gamma^{-1}(i - c_k) \right)
\end{split}\end{equation}

\end{defn}
We will consider the space $S^k_{\underline{\rho}, \underline{c}, \Gamma}$ 
of ``localized'' multilinear maps parameterized by $\theta$.  To this end, 
we let $\underline{\rho} = \{ \rho_n : n \in [0, \# \mathcal{K}], \rho_n > 0 \}$ be a sequence of radii where $\# \mathcal{K}$
is the cardinality of $\underline{c}$ (which can be infinite) and
let 
$D_{\underline{\rho}} = \{ \theta \in (\mathbb{C}^l)^{\# \mathcal{K} }/(\mathbb{Z}^l)^{\# \mathcal{K}} : 
|\text{ Im}(\theta_n)| < \rho_n \}$. 

The elements $M(\theta)$ in the space $S^k_{\underline{\rho}, \underline{c}, \Gamma}$ are multilinear maps
 on the space of localized vectors $\ell^\infty_{\underline{c}, \Gamma}$
 that depend analytically by $\theta \in D_{\underline{\rho}}$, which we assume to take the form
\begin{equation}
 M(\theta) = \sum_{n \geq 0}^{\# \mathcal{K}} M^{(n)}(\theta_1, \ldots, \theta_n)
\end{equation}
We will assume that each $M^{(n)}$ is complex differentiable in the strip $D_{\rho_n}$ and we define the norm of $M$ by
\begin{equation}
 \| M \|_{\underline{\rho}, \underline{c}, \Gamma} = \sum_{n \geq 0}^{\# \mathcal{K}}  \| M^{(n)} \|_{\rho_n, \underline{c}, \Gamma}
\end{equation}
where 
\begin{equation}
\| M^{(n)} \|_{\rho_n, \underline{c}, \Gamma} = \sup_{\theta \in D_{\rho_n}} \| M^{(n)}(\theta) \|_{\underline{c}, \Gamma}
\end{equation}
We now define the space $S^k_{\underline{\rho}, \underline{c}, \Gamma}$ by
\begin{equation}
 S^k_{\underline{\rho}, \underline{c}, \Gamma} = \{
M: D_{\underline{\rho}} \rightarrow \mathcal{L}^k_{\underline{c}, \Gamma} \text{ } : \text{ } M \in C^1_\Gamma,  
\|M(\theta) \|_{\underline{\rho}, \underline{c}, \Gamma} < \infty \}
\end{equation}

In similar spirit to Lemma \ref{complinear}, we have the following result for compositions of multilinear functions acting on the space of localized vectors.  
See \cite{FLM1} for the proof.
\begin{lemma}\label{complineartheta}
(1) If $A, B \in \mathcal{L}_{\underline{c}, \Gamma}(\ell^\infty(\mathbb{Z}^N))$ then $AB \in \mathcal{L}_{\underline{c}, \Gamma}(\ell^\infty(\mathbb{Z}^N))$ and
$$\| AB \|_{\underline{c}, \Gamma} \leq \|A\|_{\underline{c},\Gamma} \|B \|_{\underline{c}, \Gamma}$$
(2) More generally, if $A \in \mathcal{L}^k_{\underline{c}, \Gamma}(\ell^\infty(\mathbb{Z}^N))$ and
 $B_j \in \mathcal{L}^{n_j}_{\underline{c}, \Gamma}(\ell^\infty(\mathbb{Z}^N))$
for $1\leq j \leq k$.  
Then the contraction $AB_1 \cdots B_k$ defined by
$AB_1 \cdots B_k(v_1, \cdots, v_k) = A(B_1 v_1, \cdots B_k v_k)$ where $v_i \in \ell^\infty(\mathbb{Z}^N)^{n_i}$
is in
$\mathcal{L}_{\underline{c}, \Gamma}^{n_1 + \cdots + n_k}(\ell^\infty(\mathbb{Z}^N))$ and
$$\|AB_1 \cdots B_k \|_{\underline{c}, \Gamma} \leq \|A \|_{\underline{c}, \Gamma} \|B_1\|_{\underline{c}, \Gamma} \cdots \|B_k \|_{\underline{c},\Gamma}$$
(3)if $A \in S^k_{\underline{\rho},\underline{c}, \Gamma}(\ell^\infty(\mathbb{Z}^N))$ and $B_j \in S^{n_j}_{\underline{\rho},\underline{c}, \Gamma}(\ell^\infty(\mathbb{Z}^N))$
for $1\leq j \leq k$.  Then the contraction $AB_1 \cdots B_k \in S_{\underline{\rho},\underline{c}, \Gamma}^{n_1 + \cdots + n_k}(\ell^\infty(\mathbb{Z}^N))$ and
$$\|AB_1 \cdots B_k \|_{\underline{\rho},\underline{c}, \Gamma} \leq \|A \|_{\underline{\rho},\underline{c}, \Gamma} \|B_1\|_{\underline{\rho},\underline{c}, \Gamma} 
\cdots \|B_k \|_{\underline{\rho},\underline{c} \Gamma}$$
\end{lemma}
$\Box$

Now we consider the space of analytic embeddings $K:D_{\underline{\rho}} \rightarrow \mathcal{M}$ and 
 that are localized near infinitely many lattice sites.  
The reader should think of the function $K$ as giving the parameterization of the torus.
We will assume that each component $(K)_i$, $i \in \mathbb{Z}^N$ of $K$ takes the form
\begin{equation}
 (K)_i (\theta) = \sum_{n \geq 0}^{\# \mathcal{K}}  (K^{(n)})_i (\theta_1, \ldots, \theta_n)
\end{equation}
where $(K^{(n)})_i $ is a finite dimensional analytic function
 and we give a norm to $(K)_i$ by
\begin{equation}
 \| (K)_i \|_{\underline{\rho}} = \sum_{ n \geq 0}^{\# \mathcal{K}}  \|  (K^{(n)})_i \|_{\rho_n}
\end{equation}
\begin{equation}
\| K \|_{\underline{\rho}, \underline{c}, \Gamma} = \sup_{i \in \mathbb{Z}^N} \min_{k \in \mathcal{K}} \Gamma^{-1}(i - c_k) \| K_i \|_{\underline{\rho}}
 \end{equation}
We now come to the definition of the space of analytic localized embeddings of a torus, namely the space
\begin{equation}\begin{split}
& \mathcal{A}_{\underline{\rho}, \underline{c}, \Gamma}((\mathbb{T}^l)^{ \# \mathcal{K} }) = \\ &
\{ K: D_{\underline{\rho}} \rightarrow \mathcal{M}
 \text{ } : \text{ } K \text{ is analytic in } D_\rho, K \in C^0(\overline{D_\rho})
\text{, }\|K\|_{\underline{\rho}, c, \Gamma} < \infty \} 
\end{split}\end{equation}

Before defining an embedding of the stable manifold of a torus, we will need to consider the
notion of a whiskered embedding of a torus localized near
a collection $\underline{c}$ of sites, which is either finite or infinite, and ``decays
like $\Gamma$''.  Definition \ref{whiskereddef} is based on the growth and decay rates of
the cocycle generated by $DF\circ K(\theta)$, where $K$ is the 
embedding of the torus under consideration.  This is a generalization of 
the notion of a whiskered embedding in \cite{FLS}.  
In Section \ref{spec_form_defn}, we provide a notion of hyperbolicity based on the spectral properties 
of operators associated to $DK$.

\begin{defn}\label{whiskereddef}(\textbf{Growth and decay rate formulation of hyperbolicity}) Let
 $\underline{\rho} = \{ \rho_n \text{ } | \text{ } n, \rho_n \geq 0 \}$ be a sequence of radii, 
 $\underline{c} = \{ c_k \}_{k \in \mathcal{K}}$ a collection of sites indexed by $\mathcal{K}$,
 $\omega \in \left(\mathbb{R}^l \right)^{\# \mathcal{K}}$
a frequency vector,
$\Gamma$ a decay
 function and a map $F:\mathcal{M} \rightarrow \mathcal{M} \in C^1_\Gamma$.    
We say that $K: D_{\underline{\rho}} \rightarrow \mathcal{M} \in \mathcal{A}_{\underline{\rho}, \underline{c}, \Gamma}$ is a 
whiskered embedding for $F$ when we have:

\noindent 1) The tangent space has an invariant splitting
\begin{equation}\label{splitting}
T_{K(\theta)} \mathcal{M} = \mathcal{E}^s_{K(\theta)}\oplus \mathcal{E}^c_{K(\theta)} \oplus \mathcal{E}^u_{K(\theta)}
\end{equation}
where $\mathcal{E}^{s, c, u}_{K(\theta)}$ satisfy 
$DF(K(\theta)) \mathcal{E}^{s, c, u}_\theta = \mathcal{E}^{s, c, u}_{\theta + \omega}$

\noindent 2) The projections $\Pi^{s, c, u}_{K(\theta)}$ associated to this splitting are in $S^1_{\underline{\rho}, \underline{c}, \Gamma}$.

\noindent 3) The splitting \eqref{splitting} is characterized by asymptotic growth conditions:  
Let $T_\omega:D_{\underline{\rho}} \rightarrow D_{\underline{\rho}}$ be defined by $T_\omega(\theta) = \theta + \omega$ and
suppose that
there are $0 < \mu_1, \mu_2 < 1, \mu_3 > 1$ such that $\mu_1 \mu_3 < 1$, $\mu_2 \mu_3 < 1$ and $C_h > 0$ such that
for all $n \geq 1$, $\theta \in D_{\underline{\rho}}$ 
\begin{equation}\begin{split}
 \| DF \circ K \circ T^{n - 1}_\omega \times \cdots \times DF \circ K v & \|_{\underline{\rho}, \underline{c}, \Gamma} \\
& \leq C_h \mu_1^n \|v \|_{\underline{\rho}, \underline{c}, \Gamma} \iff v \in \mathcal{E}^s_{K(\theta)}.
\end{split}\end{equation}

\begin{equation}\begin{split}
 \| DF^{-1} \circ K \circ T^{n - 1}_\omega \times \cdots \times DF^{-1} \circ K v & \|_{\underline{\rho}, \underline{c}, \Gamma} \\
& \leq C_h \mu_2^n \|v \|_{\underline{\rho}, \underline{c}, \Gamma} \iff v \in \mathcal{E}^u_{K(\theta)}
\end{split}\end{equation}

\begin{equation}\begin{split}
&\| DF \circ K \circ T^{n - 1}_\omega \times \cdots \times DF \circ K v \|_{\underline{\rho}, \underline{c}, \Gamma} \leq C_h \mu_3^n \|v \|_{\underline{\rho}, \underline{c}, \Gamma} \\
&\text{and} \\
&\| DF^{-1} \circ K \circ T^{n - 1}_\omega \times \cdots \times DF^{-1} \circ K v \|_{\underline{\rho}, \underline{c}, \Gamma} \leq C_h \mu_3^n \|v \|_{\underline{\rho}, \underline{c}, \Gamma} \\ 
& \iff v \in \mathcal{E}^c_{K(\theta)}
\end{split}\end{equation}

\end{defn}
  
\begin{rmk}  The notion of a whiskered torus considered in this paper is slightly more general that the one
 considered in \cite{FLS}.  Even if the torus is finite dimensional, we can allow for $\mathcal{E}^c_{K(\theta)}$ to be infinite, which
is not done in \cite{FLS}.  Also, we work directly with infinite dimensional tori, whereas in \cite{FLS} infinite
dimensional tori are obtained by taking limits of tori of increasing dimension.  
However, the proofs of the results in this paper do not require that the map $F$ is symplectic and hence it is
 natural to consider a more general
 notion of a whiskered torus.  However, Definition \ref{whiskereddef} does have the same flavor as the whiskered tori as constructed in \cite{FLS} in
the sense the the conditions for a torus to be whiskered are on growth and decay rates. It is important to note that we also allow some of the subspaces
$\mathcal{E}^{c, s, u}$ to be empty.  

Moreover, in \cite{FLS} a quasi-Newton method is implemented to construct the whiskered tori, and as a result in \cite{FLS} it is assumed that the 
frequency vector $\omega$ is Diophantine.  In this paper, since we already assume the existence of a whiskered torus and the methods
for constructing the stable and unstalbe manifolds of a whiskered torus do not require $\omega$ to be Diophantine and hence
we make no such assumption on $\omega$.      
\end{rmk}
$\Box$

\begin{figure}
 \begin{center}
  \includegraphics[scale = 0.4]{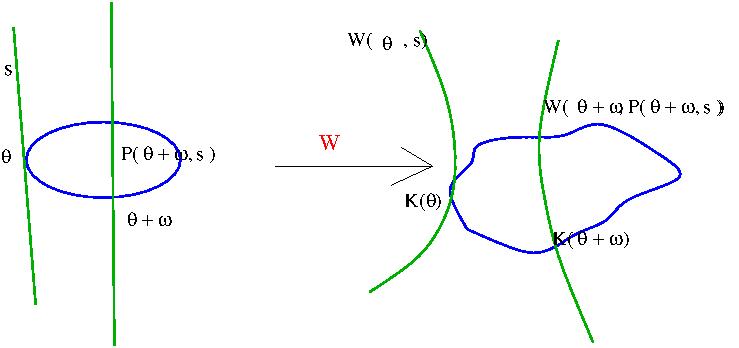}
 \end{center}
\caption{The left corresponds to the parameter space with its associated idealized dynamics modeling the actual stable
manifold for $F$ depicted on the right.  The parameterization $W$ sends the idealized parameter space into the phase 
space and preserves the canonical dynamics of rotation along the torus and contraction, given 
by a polynomial bundle map $P$, along the fibers.}
\label{fig:embed}
\end{figure}

Finally, we now describe the notion of an embedding $W$ of the stable manifold of a localized whiskered torus.  
Figure \ref{fig:embed} depicts the embedding of the stable manifold of a whiskered torus.  

We will assume that the parameterization $W$ of the stable manifold is an analytic bundle map
whose domain is the stable bundle $\mathcal{E}^s$ and range is the tangent bundle of the phase space $T \mathcal{M}$.
We recall that given two bundles $\Pi_x: X \rightarrow (\mathbb{T}^l)^{\# \mathcal{K}}$ 
and $\Pi_y: Y \rightarrow (\mathbb{T}^l)^{\# \mathcal{K}}$, a bundle map is a map
$\phi: X \rightarrow Y$ that takes fibers to fibers,
 that is if we fix $\theta$ on the torus
$\phi ( \Pi_x^{-1} (\theta) ) \subset  \Pi_y^{-1}( \phi(\theta, 0) )$.

In the specific case of the parameterization $W$, we assume that we have a 
family $W_\theta : \mathcal{E}^s_{K(\theta)} \rightarrow V(\theta)$ of maps depending on $\theta \in D_{\underline{\rho}}$.
Defining analyticity of bundle maps is done locally:  for
each $(\theta_0, s^\theta_0) \in \mathcal{E}^s$, we can trivialize the bundle
by identifying different fibers $\mathcal{E}^s_{K(\theta)}$.  In such a trivial neighborhood we can remove the dependence 
of the fibers on $\theta$ and  deal with a function $W(\theta, s)$ where $\theta$ lies in a neighborhood 
of $\theta_0$ and $s$ in a neighborhood of the origin of $E^s_{K(\theta_0)}$.  We will assume, in a trivialized neighborhood
of $(\theta_0, s^\theta_0)$ that each component $(W)_i, i \in \mathbb{Z}^N$ of $W$ takes the form

\begin{equation}\label{W_exp}
 (W)_i = \sum_{m, n \geq 0}^{m = \# \mathcal{K}, n = \text{dim}(\mathcal{E}^s_0)} W^{m,n}_i(\theta_1, \ldots, \theta_m, s_1, \ldots s_n)
\end{equation}
Where $W^{m,n}_i(\theta_1, \ldots, \theta_m, s_1, \ldots s_n)$ is an analytic function of 
finitely many variables. 

We then give each component $(W)_i$, defined globally, the sup norm
\begin{equation}
 \| (W)_i \| : = \sum_{m, n \geq 0}^{m = \# \mathcal{K}, n = \text{dim}(\mathcal{E}^s_0)} \| W^{m, n} \|_{\underline{\rho}_1, \rho_2}
\end{equation}
A norm on $W$ is then given by 
\begin{equation}
\| W  \|_{\underline{\rho}_1, \rho_2, \underline{c}, \Gamma} = \sup_{i \in \mathbb{Z}^d} \min_{k \in \mathcal{K}}\Gamma^{-1}(i - c_k)
 \| (W)_i \|_{\underline{\rho}_1, \rho_2}
\end{equation}
One question that arises is that this norm may dpened on the trivialization of the bundle $\mathcal{E}^s$.  We point out that
different trivializations lead to equivalent norms.  Of course, this contraction properties of operators may change, but we will
show that our argument applies when we consider norms associated 
in balls  of small diameter.

We will denote the space  $B_{\underline{\rho}_1, \rho_2}$ to be a neighborhood of the zero
section of the bundle $\mathcal{E}^s$, that is 
\begin{equation}
 B_{\underline{\rho}_1, \rho_2} = \{ (\theta, s_\theta) : |\theta_n| \leq \rho_n, \text{ for every } n \geq 0
 \text{ and } |s_\theta| < \rho_2 \}
\end{equation}
The space $\mathcal{A}_{\underline{\rho}_1, \rho_2, \underline{c}, \Gamma}(  B_{\underline{\rho}_1, \rho_2}, V(\theta)) $
of analytic localized bundle maps $W_\theta$ defined in a 
neighborhood $ B_{\underline{\rho}_1, \rho_2}$ of the zero section of bundle $\mathcal{E}^s$ is thus given by

\begin{equation}\begin{split}
&\mathcal{A}_{\underline{\rho}_1, \rho_2, \underline{c}, \Gamma}(B_{\rho_2}, V(\theta)) = \begin{Bmatrix} W:  B_{\underline{\rho}_1, \rho_2} \rightarrow \mathcal{M} \text { } : \text{ }
  \|W\|_{\underline{\rho}_1,\rho_2, c, \Gamma} < \infty \end{Bmatrix} 
\end{split}\end{equation}

\begin{rmk}
We chose to take the parameterization $W$ of the stable manifold of a whiskered torus to be a bundle map since
not every vector bundle over a torus is trivial, and examples of non-trivial bundles arise naturally (see \cite{FLS_finite}
for finite dimensional examples where non-trivial bundles arise naturally).

For the embeddings of the tori $K$ we can assume nonuniform radius $\rho$ for the domain of analyticity of $K^{(n)}_i$.  
The methods of this paper allow us to prove that we can choose an embedding of the stable manifold which has an
 expansion as in \eqref{W_exp} for which the domain of analyticity in the $s$ variable is uniform among $W^{(n)}$, 
which is why are definitions of the embeddings $W$ have a uniform domain in $s$.
\end{rmk}
$\Box$

\subsubsection{Spectral Formulation of of Hyperbolicity}\label{spec_form_defn}
In this section we describe a more general notion of hyperbolicity than the one in Definition \ref{whiskereddef} based
on spectral properties of operators associated to $DK$.  One can weaken the hypothesis that $K$ is
a whiskered embedding by considering the more general notion of non-resonant subspaces, similar to 
what is done in \cite{param1}.  

To this end, let $A, B: (\mathbb{T}^l)^{ \# \mathcal{K} }  \rightarrow \ell^\infty_{\underline{c}, \Gamma}$ be 
in $S^1_{\underline{\rho}, \underline{c}, \Gamma}$ and consider the
 operators $\mathcal{L}^\omega_B$, $\mathcal{R}^{k, \omega}_A$ 
acting on the space $S^k_{\underline{\rho}, \underline{c}, \Gamma}$ that are defined by
\begin{equation}\begin{split}
&(\mathcal{L}^\omega_B M)(\theta)(x_1, \ldots, x_k) = B(\theta) M(\theta - \omega)(x_1, \ldots, x_k) \\
&(\mathcal{R}^{k, \omega}_{A} M)(\theta)(x_, \ldots, x_k) = M(\theta)(x_1, \ldots, A(\theta - \omega)x_k, \ldots, x_k) \\
\end{split}\end{equation}
Instead of assuming that $K$ is a whiskered embedding in the sense of Definition \ref{whiskereddef}, one can 
replace condition $3)$ in Definition \ref{whiskereddef} by: 
\vskip 2 em 

\noindent $3^*)$  $\text{Spec}(\mathcal{L}_{A^c}) \subset \{ \mu^{-1} \leq |z| \leq \mu_3 \}$ \\
$\text{Spec}(\mathcal{L}_{A^s}) \subset \{ 0 \leq |z| \leq \mu_1 \}$ \\
$\text{Spec}(\mathcal{L}_{A^u}) \subset \{ 0 \leq |z| \leq \mu_2^{-1} \}$

\begin{rmk}
We emphasize that in definition $3^*)$ we assume the existence of the 
invariant bundles and consider the spectrum of the operators 
restricted to them.  In the papers \cite{Mather68, HaroL06} one considers
more general definitions of hyperbolicity in which one just 
assumes that  $\text{Spec}(\mathcal{L}_{A^u})$ 
has a  gap in an annulus. The main difficulty in proving the 
equivalence of this general definition with $3^*)$ is that one 
has to show 
that the spectral projections corresponding to the two components of 
$\text{Spec}(\mathcal{L}_{A})$ correspond to projections over a bundle. 
This is not too difficult to establish, but we will not consider it here.  
\end{rmk}

\begin{lemma}  
\label{rateimpliesspec} 
The rate formulation in Definition \ref{whiskereddef} 
implies the spectral formulation $3^*)$. 
\end{lemma} 

 Lemma~\ref{rateimpliesspec} is a consequence of 
the observation that if we consider the equation 
for $\beta$ given $\alpha$
\begin{equation} \label{tosolve} 
(\mathcal{L}_{A^s}) \beta  - z \beta =  \alpha 
\end{equation} 
we can obtain a solution
\begin{equation} \label{solution} 
\beta = \sum_{j= 0} z^{-j} [(\mathcal{L}_{A^s}) ]^j \alpha 
\end{equation}
Since $ [(\mathcal{L}_{A^s}) ]^j$ is 
multiplication by
$A^s(\theta + (j-1) \omega)\cdots A^s(\theta) $ and shifting, we see 
that the rate conditions imply that if $|z| > \mu_1$, 
then the series in \eqref{solution} converges absolutely. Then, one can justify the reordering of 
terms so that the series indeed is a solution. 

We also have a converse: 
\begin{lemma}\label{spectralimpliesrates} 
If a system satisfies $3^*)$ it also satisfies $3)$. 
\end{lemma}

Lemma \ref{spectralimpliesrates} follows from the Spectral Radius Formula (See Theorem 10.13 in \cite{Rudin}), which states that 
\begin{equation}\label{spec_radius}
\lim_{n \rightarrow \infty } \|(\mathcal{L}_ A)^n \|^{1/n} = \rho(\mathcal{L}_A)
\end{equation}
where  $\rho(\mathcal{A})$ is the spectral radius of $\mathcal{L}_A$. 
Indeed, the formula \eqref{spec_radius} is valid for 
any bounded linear operator in a Banach space. 

{From}
Equation \eqref{spec_radius} it follows that, for every $\epsilon > 0$
\begin{equation}\label{spen c_rad}
 \| \mathcal{L}_A^n v\| \leq C_\epsilon(\rho(\mathcal{L}_A) + \epsilon)^n \| v \|
\end{equation}
This is the desired conclusion. 
$\Box$

We emphasize that we 
have assumed that the invariant bundle exists.

\section{Statement of Results}\label{statement_section}
We state two versions of the invariant manifold theorem for localized whiskered tori, one with assumptions on growth and decay
rates, and another version based on the spectral formulation of hyperbolicity and non-resonance conditions.  

In \cite{FLS} it was shown that given an approximate whiskered embedding, a true one exists nearby.  
In \cite{FLS}, their notion of a whiskered torus is similar to Definition \ref{whiskereddef}, and hence it is
desirable to state a theorem that applies directly to the tori constructed in \cite{FLS}.  This is the content of
Theorem \ref{maintheorem}.  

We also state a more general theorem, namely an invariant manifold associated to an invariant space of the 
cocycle generated by $DF \circ K$.  We will assume that the spectrum of $DF \circ K$ restricted to this space satisfies
some ``non-resonance'' conditions (See Equations \eqref{spec_condition} and \eqref{P_lin_cond}).  These non-resonance conditions are automatically satisfied by the 
classical (strong) (un) stable manifolds, but they are also satisfied by other submanifolds.  In particular, we can make
sense of the slow manifolds in some cases.  
This is the version of the stable manifold theorem we will prove and is 
the content of Theorem \ref{maintheorem_spec}.
The proof of Theorem \ref{maintheorem_spec} follows the ideas in \cite{param1, param3}, where a proof of the stable manifold is given for fixed
 points and for
invariant manifolds in the context of general Banach spaces.  Note that, by definition, a whiskered torus is not necessarilly
 normally hyperbolic.  Indeed, in the symplectic case, there are neutral directions not tangent to the torus
 (c.f. Definition \ref{whiskereddef}). 

First, we will consider a map $F:\mathcal{M} \rightarrow \mathcal{M}$ that has a whiskered embedding of a torus $K$ and prove that
it has a stable and unstable manifold, written $W^s(K((\mathbb{T}^l)^{ \# \mathcal{K} }))$ and
$W^u(K((\mathbb{T}^l)^{ \# \mathcal{K} }))$, respectively.    The stable manifold of
$K( (\mathbb{T}^l)^{ \# \mathcal{K}})$ is characterized by the following:  for each $\theta \in (\mathbb{T}^l)^{ \# \mathcal{K} }$ there is a manifold
\begin{equation}\label{fibers}\begin{split}
 &W^s_{K(\theta)} = \{ x \in \mathcal{M} \text{ } | \text{ } d(F^n(x), F^n(K(\theta + n \omega))) \leq C_{x, \theta} \mu_1^n, n \geq 0 \} \\
 &W^u_{K(\theta)} = \{ x \in \mathcal{M} \text{ } | \text{ } d(F^n(x), F^n(K(\theta + n \omega))) \leq C_{x, \theta} \mu_1^n, n \leq 0 \}
\end{split}\end{equation}
and then $W^{s, u}(K(\mathbb{T}^d)^{\# \mathcal{K} })$ are given by
\begin{equation}\label{whole}\begin{split}
& W^s((\mathbb{T}^l)^{ \# \mathcal{K} }) = \bigcup_{\theta \in (\mathbb{T}^l)^{ \# \mathcal{K} }} W^s_{K(\theta)} \\
& W^u((\mathbb{T}^l)^{ \# \mathcal{K} }) = \bigcup_{\theta \in (\mathbb{T}^l)^{ \# \mathcal{K} }} W^u_{K(\theta)}
\end{split}\end{equation}
$ W^s_{K(\theta)}$ and $ W^u_{K(\theta)}$ are called the stable and unstable fibers of the stable and unstable manifold.  

We will prove that there is a parameterization 
$W_\theta \in \mathcal{A}_{\underline{\rho}_1, \rho_2, \underline{c}, \Gamma}(B_{\underline{\rho}_1, \rho_2}, T_{K(\theta)} \mathcal{M})$  
of the local stable manifold (by local we mean in a neighborhood
of the origin in the $s$ variable). We will construct $W$ by solving the functional equation for $W$ and $P$
\begin{equation}
F(W(\theta, s)) = W(\theta + \omega,  P(\theta,s) )
\end{equation}
where $F$ and $\omega$ are given. 
The function $P$ is a polynomial in $s$ that describes what the dynamics are in the stable direction and 
is in $\mathcal{A}_{\underline{\rho}_1, \rho_2, \underline{c}, \Gamma}(B_{\underline{\rho}_1, \rho_2}, E^s_{K(\theta)} )$

The following result is a theorem for discrete maps of the
 lattice to itself.    In Section \ref{flowsection} we show how to extend Theorem \ref{maintheorem} in the case of whiskered tori for 
flows on lattices.   

\begin{thm}\label{maintheorem}
Let $F:\mathcal{M} \rightarrow \mathcal{M}$ be a map belonging to $C^1_\Gamma(\mathcal{B})$ for any
 ball $\mathcal{B} \subset \mathcal{M}$
 and some decay function $\Gamma$
 and let $K: D_{\underline{\rho}_1} \rightarrow \mathcal{M} \in \mathcal{A}_{\underline{\rho}_1, \underline{c}, \Gamma}((\mathbb{T}^l)^{ \# \mathcal{K} })$ 
be an analytic whiskered embedding for $F$.  Suppose that $F$ has a complex analytic extension to a neighborhood of the
 torus $K((\mathbb{T}^l)^{ \# \mathcal{K}})$, i.e. there exists $\rho_2$ such that $F$ is
analytic on

\begin{equation}
\{ z \in \mathcal{M} \text{ } | \text { } |z - K(\theta)| \leq \rho_2 \text{ for some }\theta \text{ with }
|\text{ Im } (\theta_n)| < \rho_{1, n} \text{ for all } n \}
\end{equation}
Define $A(\theta):= DF(K(\theta))$ and the operators $A^{c, s, u}(\theta)$, all of which act on the space of localized vectors 
$\ell^\infty_{\underline{c}, \Gamma}$, by
\begin{equation}
A^{c, s, u}(\theta) := \Pi^{c, s, u}_{K(\theta + \omega)}DF(K(\theta))|_{\mathcal{E}^{c, s, u}_{K(\theta)}}
\end{equation}
We assume that
for some integer $L$ we have: 
%\marginpar{New definition of $L$}
\begin{equation}\label{L}
\| A^s\circ T_{L \omega} \cdots A^s  \|_{\underline{\rho}_1, \underline{c}, \Gamma} \max(1,  \| A^{-1} \|_{\underline{\rho}_1, \underline{c}, \Gamma}) < 1 
\end{equation}
where $T_\omega(\theta) =  \theta + \omega$. 

Furthermore, assume that: $A(\theta)$ is invertible for any $\theta \in D_{\underline{\rho}_1}$ with $A(\theta)^{-1}$
being uniformly bounded in $\theta \in D_{\underline{\rho}}$
Under these assumptions, 
we can find analytic maps
 $W \in \mathcal{A}_{\underline{\rho}_1, \rho_2, \underline{c}, \Gamma}(B_{\underline{\rho}_1, \rho_2}, T_{K(\theta)} \mathcal{M})$
 and $P \in \mathcal{A}_{\underline{\rho}_1, \rho_2, \underline{c}, \Gamma}(B_{\underline{\rho}_1, \rho_2}, E^s_{K(\theta)})$
 where $P$ is a polynomial in $s$.   The equation
\begin{equation}\label{stablemanifold}
F(W(\theta, s)) = W(\theta + \omega, P(\theta, s))
\end{equation}
holds in $B_{\underline{\rho}_1, \rho_2}$ and
\begin{equation}\label{W_conclusion}
\begin{array}{ll}
W(\theta,0) = K(\theta) &  
\end{array}\end{equation}
\begin{equation}\label{P_conclusion}\begin{array}{ll}
P(\theta,0) = 0 
& DP(\theta, 0) = A^s(\theta)
\end{array}\end{equation}
Finally, the stable fiber $W^s_{K(\theta)} := W( \{\theta \} \times B^s_{\rho_2})$ is the
 unique analytic invariant manifold that is 
 tangent to the linear subspace $\mathcal{E}^s_{K(\theta)}$ where $B^s_{\rho_2}$ is a neighborhood
 of the origin of $\mathcal{E}^s_{K(\theta)}$.   As a consequence 
of Equation \eqref{stablemanifold} the stable fibers
satisfy the invariance property that
\begin{equation}\label{inv_eqn_thm}
 F(W^s_{K(\theta)}) = W^s_{K(\theta + \omega)}
\end{equation}

\end{thm}

We can generalize the assumptions of Theorem \ref{maintheorem} to include non-resonant manifolds.  More 
specifically, instead of assuming that $K$ is a whiskered embedding as in Definition \ref{whiskereddef}, one can 
assume spectral properties related $DK$.

\begin{thm}\label{maintheorem_spec}
Let $F:\mathcal{M} \rightarrow \mathcal{M}$ be a map belonging to $C^1_\Gamma(\mathcal{B})$ for any
 ball $\mathcal{B} \subset \mathcal{M}$
 and some decay function $\Gamma$
 and let $K: D_{\underline{\rho}_1} \rightarrow \mathcal{M} \in \mathcal{A}_{\underline{\rho}_1, \underline{c}, \Gamma}((\mathbb{T}^l)^{ \# \mathcal{K} })$ 
be an analytic embedding for of a torus for $F$. 

Furthermore, suppose that $F$ has a complex analytic extension in a neighborhood of the
 torus $K((\mathbb{T}^l)^{ \# \mathcal{K} })$, i.e. there exists $\rho_2$ such that $F$ is
analytic on

\begin{equation}
\{ z \in \mathcal{M} \text{ } | \text { } |z - K(\theta)| \leq \rho_2 \text{ for some }\theta \text{ with }
|\text{ Im } (\theta_n)| < \rho_{1, n} \text{ for all } n \}
\end{equation}
Define $A(\theta):= DF(K(\theta))$ and the operators $A^{c, s, u}(\theta)$, all of which act on the space of localized vectors 
$\ell^\infty_{\underline{c}, \Gamma}$, by
\begin{equation}
A^{c, s, u}(\theta) := \Pi^{c, s, u}_{K(\theta + \omega)}DF(K(\theta))|_{\mathcal{E}^{c, s, u}_{K(\theta)}}
\end{equation}

We will assume that:\\
\noindent 1) $A(\theta)$ is invertible for any $\theta \in D_{\underline{\rho}_1}$

\noindent 2) The tangent space splits as 
\begin{equation}\label{splitting_spec_thm}
T_{K(\theta)} \mathcal{M} = \mathcal{E}^s_{K(\theta)}\oplus \mathcal{E}^c_{K(\theta)} \oplus \mathcal{E}^u_{K(\theta)}
\end{equation}
where $\mathcal{E}^{s, c, u}_{K(\theta)}$ satisfy:

\noindent 3) The projections $\Pi^{s, c, u}_{K(\theta)}$ associated to this splitting are in $S^1_{\underline{\rho}_1, \underline{c}, \Gamma}$.

\noindent 4)  $\text{Spec}(\mathcal{L}^\omega_{A^s}) \subset \{ z \in \mathbb{C} : |z| < 1 \}$,
where $\mathcal{L}^\omega_{A^s}$ is the transfer operator defined in Section \ref{spec_form_defn}.

\noindent 5) We have the following non-resonance condition on the transfer operators $\mathcal{L}$ and $\mathcal{R}$
\begin{equation}\label{spec_condition}
 \text{Spec}(\mathcal{L}^\omega_{ (A^{c \oplus u})}, S^i_{\underline{\rho}_1, \underline{c}, \Gamma}) \cap
\left( \text{Spec}(\mathcal{R}^{1,\omega}_{A^s}, S^i_{\underline{\rho}_1, \underline{c}, \Gamma}) \right)^i = \emptyset
\end{equation}
for $i = 1, \ldots, L$,  
where $L$ is large enough so that it satisfies

\begin{equation} \label{Lnew}  
\left( \text{Spec}(\mathcal{L}^\omega_{ (A^{c \oplus u})}, S^K_{\underline{\rho}_1, \underline{c}, \Gamma}))\right)^{-1} 
\text{Spec}(\mathcal{R}^{1,\omega}_{A^s}, S^K_{\underline{\rho}_1, \underline{c}, \Gamma}))^K 
\subset \{ z \in \mathbb{C} | |z| < 1\}. 
\end{equation} 
for all $K > L$.

Then we have conclusions \eqref{W_conclusion} and \eqref{P_conclusion} of Theorem \ref{maintheorem}.  Moreover, if we suppose that
\begin{equation}\label{P_lin_cond}
  (\text{Spec}(\mathcal{R}^\omega_{A^s} , S^i_{\underline{\rho}_1, \underline{c}, \Gamma}))^i \cap 
  \text{Spec}(\mathcal{L}^\omega_{A^s}, S^i_{\underline{\rho}_1, \underline{c}, \Gamma}) = \emptyset
\end{equation}
for $1 \leq i \leq L$, then we can chose $P$ to be linear.  

Finally, the stable fiber $W^s_{K(\theta)} := W( \{\theta \} \times B^s_{\rho_2})$ is the
 unique analytic invariant manifold that is 
 tangent to the linear subspace $\mathcal{E}^s_{K(\theta)}$ and as a consequence 
of Equation \eqref{stablemanifold} the stable fibers
satisfy the invariance property that
\begin{equation}
 F(W^s_{K(\theta)}) = W^s_{K(\theta + \omega)}
\end{equation}

\end{thm}

%\marginpar{This remark is new} 
\begin{rmk} 
The number $L$ introduced in 5), can always be found
 because the spectral properties assumed of 
$\mathcal{L}^\omega_{A^s}$ imply that the norm of the shifted 
product goes to $0$ as $k$ goes to infinity. Hence, we can find a
value $L$ such that it remains below $1$. 

\end{rmk}

\begin{rmk}
Note that if $K$ is a whiskered embedding for $F$, then $K$ satisfies
assumptions $2$ - $5$.    
We will prove Theorem \ref{maintheorem_spec}, which generalizes Theorem \ref{maintheorem}, in Section \ref{proof}.
Theorem \ref{maintheorem} is the most natural theorem to state 
on the existence of localized stable manifolds for the whiskered tori constructed in \cite{FLS}.
Theorem \ref{maintheorem_spec} yields also non-resonant manifolds, slow
manifolds, and gives conditions that allow to choose $P$ linear, namely condition \ref{P_lin_cond}.

Moreover, the fact that $P$ satisfies $DP(\theta, 0) = A^s(\theta)$ implies 
that $W(B_{\underline{\rho}_1, \rho_2})$ satisfies Equation \eqref{fibers}.  

Although the statement of  Theorem \ref{maintheorem_spec} only gives the stable manifold for a whiskered embedding,
 we can use Theorem \ref{maintheorem_spec}
to construct the unstable manifold $W^u((\mathbb{T}^l)^{ \# \mathcal{K} })$ by noting that the unstable manifold
for the torus $K((\mathbb{T}^l)^{ \# \mathcal{K} })$ under the map
 $F$ is simply the stable manifold for $K((\mathbb{T}^l)^{ \# \mathcal{K} })$ under the map $F^{-1}$.  

The stable fibers $W^s_{K(\theta)}$ also satisfy the usual graph property, namely that
$W^s_{K(\theta)}$ is, in a neighborhood of $K(\theta)$, a graph over $\mathcal{E}^s_{K(\theta)}$.  
Indeed, since Theorem \ref{maintheorem_spec} implies that  

\begin{equation}\label{tangencyproperty}
T_{K(\theta)} W^s_{K(\theta)} =  \mathcal{E}^s_{K(\theta)}
\end{equation}
and hence if we write $W = (W^s, W^{c \oplus u})$, where $W^s = \Pi^s_{K(\theta)} W$ 
and $W^{c \oplus u}_{K(\theta)} = \Pi^{c \oplus u}_{K(\theta)} W$, then 
\eqref{tangencyproperty} implies that $D_s W^s(\theta, 0)$ is invertible and hence by the
implicit function theorem $W^s(\theta, s)$ is invertible in $s$ in a neighborhood of $s = 0$.
Thus, if $H_\theta(s) := W^{c \oplus u} \circ (W^s)^{-1}$ then the point $(s, H(s))$ is
in the image of $W$, which means that $W^s_{K(\theta)}$ is the graph of $H_\theta$.
One of the main conclusions of Theorem \ref{maintheorem_spec} is that the function whose graph is the stable manifold
has decay properties.  

\end{rmk}

\subsection{Non-uniqueness of $(W, P)$}\label{uniqueness_statement_section}
The parameterization $W$ and the function $P$ of Theorems \ref{maintheorem} and \ref{maintheorem_spec}
 are not unique as the theorems are stated, though the image of $W$ is unique.  The origin of non-uniqueness 
 comes from the fact that if the pair $(W, P)$ satisfies the conclusions of Theorem \ref{maintheorem_spec}
then a change of coordinates in the stable direction will yield another solution.  More precisely,
if $Q_\theta : \mathcal{E}^s_\theta \rightarrow \mathcal{E}^s_\theta$ is a polynomial bundle map satisfying $Q_\theta(0) = 0$ and $DQ_\theta(0) = \text{Id}$ and we
let $\tilde{W}_\theta = W_\theta \circ Q_\theta$ and $\tilde{P}_\theta = 
\left( Q_{\theta + \omega}^{-1} \circ P_\theta \circ Q_{\theta} \right)^{\leq L}$ where 
$\left( Q_{\theta + \omega}^{-1} \circ P_\theta \circ Q_{\theta} \right)^{\leq L}$ denotes the truncation of
$Q_{\theta + \omega}^{-1} \circ P_\theta \circ Q_{\theta}$ up to order $L$ then the pair $(\tilde{W}, \tilde{P})$ is
also satisfies the conclusions of Theorem \ref{maintheorem_spec}.  

The following Lemma states that if one specifies certain conditions on the first order term of $P$ and
the first $L$ order terms on $W$ determine uniquely the higher order terms of $W$ and $P$.

 \begin{lemma}\label{uniqueness_lem}
 Under the setup of Theorem \ref{maintheorem} or Theorem \ref{maintheorem_spec} the fibers 
$W(\{\theta\} \times B^s_{\rho_2})$ are unique in the sense that any localized analytic invariant manifold 
tangent to $\mathcal{E}^s_{K(\theta)}$ coincides with $W(\{\theta\} \times B^s_{\rho_2})$ in a neighborhood of
$K(\theta)$.  Moreover, suppose that we have polynomial bundle maps $W^{\leq L}$ and $P$ 
of degree $L$ in $s$ that satisfy
\begin{equation}
 F \circ W^{\leq L}(\theta, s) = W^{\leq L}(\theta + \omega, P(\theta, s)) + o(|s|^L)
\end{equation}
Then there is a unique $W^>$ 
such that the pair $(W^{\leq L} + W^>, P)$ satisfies the conclusion of Theorem \ref{maintheorem_spec}.  Thus,
if we specify the solution up to order $L$, the higher order terms are unique.  The 
non-uniqueness of the low-order terms can be classified as follows:
Suppose the following:

\noindent 1) $W(\theta, 0) = K(\theta)$

\noindent 2)  $DW(\theta, 0)$ is specified

\noindent 3) $P(\theta, 0) = 0$ and $DP(\theta, 0) = A^s(\theta)$

\noindent 4) $W_{i, \theta}^s = 0$, where $W_{i, \theta}^s := \Pi^s_{K(\theta)} D^i_s W(\theta, 0)$  

\noindent Then the parameterization $W$ and the polynomial $P$ are unique.
\end{lemma}
As mentioned  in Section \ref{uniqueness_section}, Lemma \ref{uniqueness_lem}  is a direct consequence of our proof of Theorem \ref{maintheorem_spec}.
 A consequence of Lemma \ref{uniqueness_lem} is the following lemma
 that reduces proving Theorem \ref{maintheorem_spec} for the map $F^n$ instead of $F$, which is useful
 since we have better estimates on $F^n$ compared to $F$.  Moreover, the ideas of the proof
 of Lemma \ref{F^n_vs_F} are used to prove the result for flows in Section \ref{flowsection}.  
  
\begin{lemma}\label{F^n_vs_F}
 Suppose that, for the map $F^n$, the pair
 $(W, P)$ satisfy the conclusions of Theorem \ref{maintheorem_spec}
 and is the unique pair satisfying the hypotheses of Lemma \ref{uniqueness_lem}.  
 Then there is a polynomial in the variable $s$ $R$ such that the pair $(W, R)$ satisfies the conclusions of
 Theorem \ref{maintheorem_spec} for the map $F$, in particular we have that
 \begin{equation}\label{R_eqn}
  F \circ W(\theta, s) = W(\theta + \omega, R(\theta, s))
 \end{equation}
\end{lemma}
\textit{Proof:}
We write 
\begin{equation}
 R(\theta, s) = \sum_{i = 1}^L R_{\theta, i} (s, \ldots, s)
\end{equation}
and then solve for each $R_{\theta, i}$ by matching powers.  For the zeroth and first order terms we 
take $R(\theta, 0) = 0$ and $DR(\theta, 0) = A^s(\theta)$.

Using the fact that $R_\theta := R(\theta, \cdot)$
is invertible in a neighborhood of $s = 0$ we can rewrite Equation \eqref{R_eqn} to solve for the higher order terms
\begin{equation}\label{R_inv_eqn}
 F \circ W(\theta, R_{\theta}^{-1}(s)) = W(\theta + \omega ,s)
\end{equation}
This form is more convenient since we will exploit the uniqueness result of Lemma \ref{uniqueness_lem}
to find $R$ such that the left satisfies the hypotheses of Lemma \ref{uniqueness_lem}.  
Note that $R^{-1}_\theta : \mathcal{E}^s_{\theta + \omega} \rightarrow \mathcal{E}^s_\theta$
is a bundle map.  

Taking $i$ derivatives of Equation \eqref{R_inv_eqn} for $i = 2, \ldots L$ we obtain
\begin{equation}\label{R2_eqn}
 - DF \circ W_{\theta, 1} \left(A^s_{\theta + \omega} \right)^{-1} R_{\theta, i} ((A^s(\theta) )^{-1})^{\otimes i}
 + r_{\theta, i} = W_{\theta, i}
\end{equation}
Where $r_{\theta, i}$ is a term involving $F$, $W$, their derivatives,
and derivatives of $R$ of order $i - 1$ or smaller.  We have
also used the chain rule to relate $D^i \left( R_{\theta} ^{-1} \right)$
to $D^i R_{\theta}$
%We
% have used the general fact that $f \circ g = \text{Id}$ implies 
%$$D^2 f = - Df (D^2 g)\circ f \left(  \left( (Dg)\circ f \right) ^{-1}\right)^{\otimes 2}$$.  
If we project Equation \eqref{R2_eqn} on the stable subspace we obtain
\begin{equation}
 -A^s(\theta + \omega) W_{\theta, 1} \left( A^s_{\theta + \omega} \right)^{-1} R_{\theta, i} ((A^s(\theta) )^{-1})^{\otimes i}
  = -\Pi^s_\theta r_{\theta, i}
\end{equation}

Since all the terms on the left not involving $R_{\theta, i}$ are invertible, it follows that
we can solve for $R_{\theta, i}$.  We then conclude
that $F \circ W(\theta, R_{\theta}^{-1}(s))$ satisfies the uniqueness assumptions that $W$ also satisfies, 
and thus $F \circ W(\theta, s) = W(\theta + \omega, R_\theta (s))$

$\Box$

\subsection{Stable Manifolds for Flows}\label{flowsection}
In this section we will extend Theorem \ref{maintheorem} to the case of flows for lattice systems.

In \cite{FLS} it was proven that a decay
vector field $X \in C^r_\Gamma(\mathcal{B})$ generates a flow $ \{ S_t \}_{t \in \mathbb{R}}$ such that $S_t$ is a decay diffeomorphism for all $t$.
More precisely, we have  

\begin{prop}\label{existenceodes}
 Let $X$ be a $C^1$ vector field on an open subset $\mathcal{B} \subset \mathcal{M}$ and consider the differential equation
\begin{equation}\label{ode}
 \dot{x}  = X(x)
\end{equation}
Let $\mathcal{B}_1 \subset \mathcal{B}$ be an open set such that $d(\mathcal{B}_1, \mathcal{B}^c) = \eta > 0$.  

Then there exist $T > 0$ such that for all initial conditions $x_0 \in \mathcal{B}_1$ there is a unique solution $S_t$ of the Cauchy problem corresponding
to Equation \eqref{ode} defined for $|t| < T$.  We denote by $S_t(x_0) = x_t$.  By uniqueness, we have that
$S_{t + s} = S_t \circ S_s$ when all the maps are defined and the composition makes sense.  Moreover

(1)For all $t \in (-T, T), S_t:\mathcal{B}_1 \rightarrow \mathcal{B}$ is a diffeomorphism onto its image.
(2) If $X \in C^1_\Gamma(\mathcal{B}_1)$ then $S_t \in C^1_\Gamma(\mathcal{B}_1)$ for all $t \in (-T, T)$.  Moreover, there exist $C, \mu > 0$ such that
\begin{equation}
 \|D S_t(x) \|_\Gamma \leq C e^{\mu t}
\end{equation}
for $x \in \mathcal{B}_1$ and $t \in (-T, T)$.  When $\mathcal{B} = \mathcal{M}$ we have $T = \infty$.  
\end{prop}

Note that $C^1_\Gamma$ functions are uniformly bounded, which is important to point out
especially in the case $\mathcal{B}$ is unbounded, e.g. when $\mathcal{B} = \mathcal{M}$.  Without the assumption that
the vector field is uniformly bounded, we would not be able to chose $T = \infty$ in the case $\mathcal{B} = \mathcal{M}$.

 Now that we have Proposition \ref{existenceodes}, we explain how to extend Theorem
\ref{maintheorem} in the case for flows.  Let $X$ be a analytic vector field with decay in $C^1_\Gamma(\mathcal{B}_1)$.  
The notion of a whiskered torus for a flow is that we have an analytic 
embedding $K: D_{\underline{\rho}} \rightarrow \ell^\infty(\mathbb{Z}^N)$ in
$\mathcal{A}_{\underline{\rho}, \underline{c}, \Gamma}$ such that 
\begin{equation}\label{floweqnflow}
 S_t \circ K(\theta) = K(\theta + t \omega)
\end{equation}
Or, equivalently, if one takes the derivative of Equation \eqref{floweqnflow} with respect to $t$ at $t = 0$ one 
obtains an equivalent equation in terms of the vector field $X$
\begin{equation}\label{floweqnvector}
 X \circ K(\theta) = \partial_\omega K(\theta) := DK(\theta) \omega
\end{equation}
which is the equation that is solved in \cite{FLS}.  We now state the definition of a localized whiskered torus
for the flow $S_t$.  

\begin{defn}\label{whiskereddef_flow}(\textbf{Whiskered tori for flows}) Let
 $\underline{\rho} = \{ \rho_n \text{ } | \text{ } n, \rho_n \geq 0 \}$ be a sequence of radii, $\omega \in \mathbb{R}^\infty$
a frequency vector, 
$\underline{c} = \{ c_n \in \mathbb{Z}^n \text{ } | \text{ } n \geq 0 \}$ a collection of lattice sites,
$\Gamma$ a decay
 function and a vector field $X$ that is in $C^1_\Gamma(\mathcal{B})$ for every ball $\mathcal{B}$ of $\mathcal{M}$.
Suppose that the flow $S_t$ exists for all time $t$.        
We say that $K: D_{\underline{\rho}} \rightarrow \mathcal{M} \in \mathcal{A}_{\underline{\rho}, \underline{c}, \Gamma}$ is a 
whiskered embedding for the flow  $S_t$ when we have:

\noindent 1) The tangent has an invariant splitting 
\begin{equation}\label{splitting_flow}
T_{K(\theta)} \mathcal{M} = \mathcal{E}^s_{K(\theta)}\oplus \mathcal{E}^c_{K(\theta)} \oplus \mathcal{E}^u_{K(\theta)}
\end{equation}
where $\mathcal{E}^{s, c, u}_{K(\theta)}$ satisfy 
$DK(\theta) \mathcal{E}_\theta^{s, c, u} = \mathcal{E}_{\theta + t\omega}^{s, c, u}$.  Moreover, we also assume

\noindent 2) The projections $\Pi^{s, c, u}_{K(\theta)}$ associated to this splitting are in $S^1_{\underline{\rho}, \underline{c}, \Gamma}$.

\noindent 3) The splitting \eqref{splitting} is characterized by asymptotic growth conditions: 
Define $A(\theta, t) := DS_t( K(\theta))$ and 
\begin{equation}
A^{s, c, u}(\theta, t) := A (\theta, t)|_{\mathcal{E}^{s, c, u}_\theta}
\end{equation}
We assume that there are $0 < \mu_1, \mu_2 < 1, \mu_3 > 1$ such that $\mu_1 \mu_3 < 1$, $\mu_2 \mu_3 < 1$ and $C_h > 0$ such that
for all $n \geq 1$ $\theta \in D_{\underline{\rho}}$ 

\begin{equation}\begin{split}
& \|A^s(\theta, t) A^s(\theta, t_0)^{-1}  \|_{\underline{\rho}_1, \rho_2, \underline{c}, \Gamma} \leq C_h e^{-\mu_1(t - t_0)}  \text{     for } t  > t_0 \geq 0\\
& \|A^u(\theta, t) A^u(\theta, t_0)^{-1} \|_{\underline{\rho}_1, \rho_2, \underline{c}, \Gamma} \leq C_h e^{\mu_2(t - t_0)}  \text{     for } t  < t_0 \leq 0 \\
& \|A^c(\theta, t) A^c(\theta, t_0)^{-1} \|_{\underline{\rho}_1, \rho_2, \underline{c}, \Gamma} \leq C_h e^{\mu_3|t - t_0|}  \text{     for all } t _0, t
\end{split}\end{equation}  
\end{defn}

We now state the analogous invariant manifold theorem in the case of flows, which is a straight-forward consequence of
Theorem \ref{maintheorem}.  

\begin{thm}\label{maintheorem_flows}
Let $X:\mathcal{M} \rightarrow \mathcal{M}$ be a vector field belonging to $C^1_\Gamma(\mathcal{B})$ for any
 ball $\mathcal{B} \subset \mathcal{M}$ and
 and some decay function $\Gamma$.  By Proposition \ref{existenceodes} the flow $S_t$ exists and is a decay
diffeomorphism for an interval $(-T. T)$, we will assume that $T = \infty$.   
 
Suppose that $K: D_{\underline{\rho}_1} \rightarrow \mathcal{M} \in \mathcal{A}_{\underline{\rho}_1, \underline{c}, \Gamma}((\mathbb{T}^l)^{ \# \mathcal{K} })$ 
is an analytic whiskered embedding for $S_t$.  Suppose that $X$ has a complex analytic extension in a neighborhood of the
 torus $K((\mathbb{T}^l)^{ \# \mathcal{K} })$, i.e. there exists $\rho_2$ such that $X$ is
analytic on

\begin{equation}
\{ z \in \mathcal{M} \text{ } | \text { } |z - K(\theta)| \leq \rho_2 \text{ for some }\theta \text{ with }
|\text{ Im } (\theta_n)| < \rho_{1, n} \text{ for all } n \}
\end{equation}
Define $A(\theta, t):= D S_t(K(\theta))$ and the operators $A^{c, s, u}(\theta, t)$, all of which act on the space of localized vectors 
$\ell^\infty_{\underline{c}, \Gamma}$, by
\begin{equation}
A^{c, s, u}(\theta, t) := \Pi^{c, s, u}_{K(\theta + \omega)}DS_t(K(\theta))|_{\mathcal{E}^{c, s, u}_{K(\theta)}}
\end{equation}
Since the embedding is whiskered, we know that
for some integer $L$ and time $t$
\begin{equation}\label{L_flow}
\| A^s(\cdot, t) \|_{\underline{\rho}_1, \underline{c}, \Gamma}^{L + 1} \| A^{-1}(\cdot, t) \|_{\underline{\rho}_1, \underline{c}, \Gamma} < 1 
\end{equation}
We will assume that:  $A(\theta, t)$ is invertible for any $\theta \in D_{\underline{\rho}_1}$
and the norm of $A^{-1}(\theta, t)$ is uniformly controlled in $\theta$.  
Under these assumptions, 
we can find analytic maps
 $W \in \mathcal{A}_{\underline{\rho}_1, \rho_2, \underline{c}, \Gamma}(B_{\underline{\rho}_1, \rho_2}, T_{K(\theta)} \mathcal{M})$
 and $P \in \mathcal{A}_{\underline{\rho}_1, \rho_2, \underline{c}, \Gamma}(B_{\underline{\rho}_1, \rho_2}, E^s_{K(\theta)})$.
 The equation
 \begin{equation}\label{stablemanifold_flows}
  S_t (W(\theta, s)) = W(\theta + t \omega, P(\theta, s))
 \end{equation}
holds in $B_{\underline{\rho}_1, \rho_2}$ and
\begin{equation}\begin{array}{ll}
W(\theta,0) = K(\theta) & 
\end{array}\end{equation}
\begin{equation}\begin{array}{ll}
P(\theta,0) = 0 
& DP(\theta, 0) = A^s(\theta, t)
\end{array}\end{equation}
Finally, the stable fiber $W^s_{K(\theta)} := W( \{\theta \} \times B^s_{\rho_2})$ is the
 unique analytic invariant manifold that is 
 tangent to the linear subspace $\mathcal{E}^s_{K(\theta)}$ and as a consequence 
of Equation \eqref{stablemanifold_flows} the stable fibers
satisfy the invariance property that
\begin{equation}
 S_t(W^s_{K(\theta)}) = W^s_{K(\theta + t\omega)}
\end{equation} 

\end{thm}
\begin{rmk}
 The proof of Theorem \ref{maintheorem_flows} only requires that $S_{t_0}$ satisfies the conditions of Theorem
\ref{maintheorem_spec} for a time $t_0$.  We chose to state \ref{maintheorem_flows} the way we did since 
it uses the same notion of a whiskered torus in the case of flows for lattice systems with 
localized interactions considered in \cite{FLS}.  
\end{rmk}

\textit{Proof:}
%Our assumptions imply that $K$ is a whiskered embedding for $S_{t_0}$ for any $t_0$.  Fix $t_0 \in \mathbb{R}$
%and denote by
%$W^s$ the stable manifold for the torus constructed in Theorem \ref{maintheorem}, and let
% $\{W^s_{K(\theta)} \text{ } | \text{ } \theta \in (\mathbb{T}^l)^\mathbb{N} \}$
% be the corresponding stable fibers, that is $W^s_{K(\theta)} = W(\{ \theta \}, B_{\rho_2})$ where $W$ is the parameterization
%constructed in Theorem \ref{maintheorem}.  

%We wish to show that $W^s$ is in fact the stable manifold of $K((\mathbb{T}^l)^\mathbb{N})$ for the flow $S_t$. 
Our assumptions imply that $K$ is a whiskered embedding for $S_{t_0}$ for some $t_0$.  
Let $(W, \tilde{P})$ satisfy the conclusion of Theorem \ref{maintheorem_spec} for the map $S_{t_0}$.  In particular we have
\begin{equation}\label{S_t0_eqn}
 S_{t_0} \circ W(\theta, s) = W(\theta + t_0 \omega, \tilde{P}(\theta , s))
\end{equation}
Applying $S_t$ to Equation \eqref{S_t0_eqn} gives
\begin{equation}
 S_{t_0} (S_t \circ W(\theta, s)) = (S_t \circ W)(\theta + t_0 \omega, \tilde{P}(\theta , s))  
\end{equation}
By the same argument given in Lemma \ref{F^n_vs_F} we can find a polynomial $P$ in $s$ such that
\begin{equation}
 S_t \circ W(\theta, s) = W(\theta + t \omega, P(\theta, s))
\end{equation}

$\Box$

\section{Proof of Theorem 2}\label{proof}
We will prove Theorem \ref{maintheorem_spec}  for $F^n$ for fixed $n > 0$ instead of $F$ itself.  The reason we do this is because we have better estimates
on $DF^n$.  For example, we are assuming that 
\begin{equation}
 \text{Spec}(\mathcal{L}^\omega_{A^s}) \subset \{ z \in \mathbb{C} : |z| < 1 \}
\end{equation}
And so by the Spectral Radius Formula
\begin{equation}
 \| \left( \mathcal{L}^\omega_{A^s} \right)^n \|_{\underline{\rho}, \underline{c}, \Gamma} 
 \leq C_\epsilon ( \rho(\mathcal{L}^\omega_{A^s}) + \epsilon )^n 
\end{equation}
However, when comparing the map $F^n$ and $F$ note that
\begin{equation}
 \mathcal{L}^\omega_{A^s} (F^n)  =  \left( \mathcal{L}^\omega_{A^s}(F) \right)^n
\end{equation}
Hence it follows that $ \mathcal{L}^\omega_{A^s} (F^n) $, and therefore $A^s(F^n)$, is a 
contraction. 

Moreover, the stable manifold of
the whiskered torus $K$ is the same for the map $F^n$ and $F$, that is
\begin{equation}\label{f^n}
W^s_{K(\theta)}(F^n) = W^s_{K(\theta + n\omega)}(F)
\end{equation}
Equation \eqref{f^n}
is a consequence of Corollary \eqref{F^n_vs_F} in Section \ref{uniqueness_statement_section}.  

We will write the solution as
\begin{equation}\label{ansatz}\begin{split}
&W(\theta, s) =  W^{\leq} + W^> = \sum_{i =0}^L W_{\theta, i}(s, \ldots, s) + W^> \\
&P(\theta, s)  = \sum_{i = 0}^L P_{\theta, i}(s, \ldots, s) 
\end{split}\end{equation}

Where $W_{\theta, i}, P_{\theta, i}$ are homogeneous polynomials in $s$, which is to say that they are $i-$multi-linear functions in $s$ and we also 
assume $W^>$ vanishes up to order $L$ in $s$.

\subsection{Finding the Low-order Terms}
We first find $W^\leq$ and $P$ by matching powers in this section and then in Sections 
\ref{fixed_section1} and \ref{fixed_section2} we find $W^>$ using a fixed point argument.

\begin{prop}\label{induction}
Assuming the hypotheses of Theorem \ref{maintheorem}, then\\
we can find  polynomials in $s$ 
\begin{equation}\begin{split}
&W^\leq = \sum_{i = 1}^L W_{i, \theta}(s, \ldots, s)\\
&P = \sum_{i = 1}^L P_{i, \theta}(s, \ldots, s)
\end{split}\end{equation}
Where $W_{i, \theta}, P_{i, \theta}$ are homogeneous polynomials of degree $i$ in $S^i_{\underline{\rho}_1, \underline{c}, \Gamma}$.
$W^\leq$ and $P$ are of degree not larger than $L$, are in 
 $ \mathcal{A}_{\underline{\rho}_1, \rho_2, \underline{c}, \Gamma}(B_{\underline{\rho}_1, \rho_2}, T_{K(\theta)} \mathcal{M})$
 and $\mathcal{A}_{\underline{\rho}_1, \rho_2, \underline{c}, \Gamma}(B_{\rho_2}, E^s_{K(\theta)})$, respectively
 for any $\rho_2 > 0$ and $\underline{\rho}_1$
satisfying $K \in \mathcal{A}_{\underline{\rho}_1, \underline{c}, \Gamma}$ and 
\begin{equation}\label{inductivesolution}
F(W^\leq(\theta, s)) = W^\leq(\theta + \omega, P^\leq( \theta, s)) + o(|s|^L)
\end{equation}
Finally, we also have that
\begin{equation}\begin{array}{ll}
W(\theta,0) = K(\theta) &  
\end{array}\end{equation}
\begin{equation}\begin{array}{ll}
P(\theta,0) = 0 
& DP(\theta, 0) = A^s(\theta)
\end{array}\end{equation}
\end{prop}
\noindent To prove Proposition \ref{induction} we will use the following

\begin{lemma}\label{technicalspectrumlemma}
Let $A, B: (\mathbb{T}^l)^{\# \mathcal{K} } \times \ell^\infty_{\underline{c}, \Gamma} \rightarrow \ell^\infty_{\underline{c}, \Gamma}$ be in $S^1_{\underline{\rho}, \underline{c}, \Gamma}$ and consider the
 operators $\mathcal{L}^\omega_B$, $\mathcal{R}^k_A$ and
$\mathcal{L}^\omega_{k, A, B}$ acting on the space $S^k_{\underline{\rho}, \underline{c}, \Gamma}$ that are defined by
\begin{equation}\begin{split}
&(\mathcal{L}^\omega_B M)(\theta)(x_1, \ldots, x_k) = B(\theta) M(\theta - \omega)(x_1, \ldots, x_k) \\
&(\mathcal{R}^{k, \omega}_{A} M)(\theta)(x_, \ldots, x_k) = M(\theta)(x_1, \ldots, A(\theta - \omega)x_k, \ldots, x_k) \\
& (\mathcal{L}^\omega_{k, A, B} M)(\theta)(x_1, \ldots, x_k)  = \\ & B(\theta) M(\theta - \omega)(A( \theta - 2\omega)x_1, \ldots, A(\theta - 2\omega )x_k) 
\end{split}\end{equation}

We have the following spectral inclusion
\begin{equation}\label{specinclusion}\begin{split}
&\text{Spec}(\mathcal{L}^\omega_{k, A, B}, S^k_{\underline{\rho}, \underline{c}, \Gamma}) \subset \text{Spec}(\mathcal{L}^\omega_B, S^k_{\underline{\rho}, \underline{c}, \Gamma})  
\text{Spec}(\mathcal{R}^{1, \omega}_A, S^k_{\underline{\rho}, \underline{c}, \Gamma})^k
\end{split}\end{equation}
Moreover, we also have that $1 \notin$ Spec$(\mathcal{L}^\omega_{k, A^s, (A^{c \oplus u})^{-1}})$.

\end{lemma}
\begin{rmk} Lemma \ref{technicalspectrumlemma} is an important way in which our proof differs from \cite{param1}.  In the present work,
Lemma \ref{technicalspectrumlemma} states that we are able
to deduce spectral properties about $\mathcal{L}^\omega_{k, A, B}$ knowing the spectral
properties of $\mathcal{L}^\omega_{B}$ and $\mathcal{R}^{1, \omega}_A$. 

In contrast, since the main theorems in \cite{param1} are stated for fixed points and normally hyperbolic invariant manifolds
 and not whiskered tori, the analogue of Lemma \ref{technicalspectrumlemma} used in \cite{param1} is easier to state.  
 Indeed, Proposition $3.2$ in \cite{param1}
relates the spectrum of certain operators $\mathcal{L}_B$, $\mathcal{L}_{k, A, B}$ and $\mathcal{R}^k_A$ in terms of
the spectrum of $A$ and $B$ directly.  More specifically, the operators considered in \cite{param1} do not depend on $\theta$, and 
this allows \cite{param1} to prove Spec$(\mathcal{L}_{k, A, B}) \subset \text{ Spec}(B) (\text{ Spec}(A))^k$.  In our case one cannot directly
relate the spectrum of $\mathcal{L}^\omega_{k, A, B}$ to the spectrum of $A$ and $B$.  Nevertheless, 
Lemma \ref{technicalspectrumlemma} in the
present paper is sufficient for our purposes since the crucial property that is needed to prove an inductive result such as
Proposition \ref{induction} both in this paper and in \cite{param1} is that
$1 \notin$ Spec$(\mathcal{L}^\omega_{k, A^s, (A^{c \oplus u})^{-1}})$.  

\end{rmk}

\textit{Proof of Lemma \ref{technicalspectrumlemma}:} 
The fact that the range of $\mathcal{L}^\omega_B, \mathcal{R}^{k, \omega}_B, \mathcal{L}^\omega_{k, A, B}$ lies in $S^k_{\underline{\rho}_1, \underline{c}, \Gamma}$
 follows from 
Lemma \ref{complineartheta}.  
Notice that $\mathcal{L}^\omega_{n, A, B} = \mathcal{L}^\omega_B \mathcal{R}^{1, \omega}_A \cdots \mathcal{R}^{k, \omega}_A$, and moreover the 
operators $\mathcal{L}^\omega_B, \mathcal{R}^{k, \omega}_A$ commute. 
 Moreover, $\text{Spec}(\mathcal{R}^{k, \omega}_A) = \text{Spec}(\mathcal{R}^{1, \omega}_A)$.  Hence using the general fact that
 $\text{Spec}(AB) \subset \text{Spec}(A) \text{Spec}(B)$  for any commuting elements of a Banach algebra \cite{Rudin}, we obtain
$$\text{Spec}(\mathcal{L}^\omega_{k, A, B}, S^k_{\underline{\rho}, \underline{c}, \Gamma}) \subset \text{Spec}(\mathcal{L}^\omega_B, S^k_{\rho, \underline{c}, \Gamma})  
\text{Spec}(\mathcal{R}^{1, \omega}_A, S^k_{\underline{\rho}, \underline{c}, \Gamma})^k$$

Finally, to show that $1 \notin$ Spec$(\mathcal{L}^\omega_{k, A^s, (A^{c \oplus u})^{-1}})$,  it suffices to show, by 
\eqref{specinclusion} that
\begin{equation}\label{specnotequal}
 \lambda^{c \oplus u} \neq \lambda_1^s \cdots \lambda_k^s
\end{equation}
where $\lambda^{c \oplus u} \in \text{Spec}(\mathcal{L}^\omega_{(A^{c \oplus u})^{-1}}, S^k_{\underline{\rho}, \underline{c}, \Gamma})$
and $\lambda_i^s \in \text{Spec}(\mathcal{R}^{1, \omega}_{A^s}, S^k_{\underline{\rho}, \underline{c}, \Gamma}), i = 1, \ldots, k$.  
Since we are working with $F^n$ and not $F$, we have $ \lambda_i^s \leq  C_h (\mu_1)^n$ and $\frac{1}{2C_h(\mu_3)^n} \leq \lambda^{c \oplus u}$
Thus if $ \lambda^{c \oplus u} = \lambda_1^s \cdots \lambda_k^s$, then $ \frac{1}{2C_h(\mu_3)^n} \leq C_h^k (\mu_1)^{kn}
$, that is
\begin{equation}
 \frac{1}{2C_h^{k + 1}} \leq  (\mu_3 \mu_1^k)^{n}
\end{equation}
However, $(\mu_3 \mu_1^k)^{n} \rightarrow 0$ as $n \rightarrow \infty$ since $\mu_1 \mu_3 < 1$, which is a contradiction. 
$\Box$

\textit{Proof of Proposition \ref{induction}:} That we can solve for the $i  = 0, 1$ terms is, as we now explain, a consequence of
our assumption that $K$ is a whiskered embedding.  More precisely, to solve for 
$W_{\theta, 0}, P_{ \theta, 0}$ we substitute \eqref{ansatz} into \eqref{stablemanifold} and evaluate at $s = 0$, and we obtain
\begin{equation}\label{zero}
F \circ W_{\theta, 0} = W_{ \theta + \omega, 0} \circ P_{ \theta, 0}
\end{equation}
Which is solved by taking $W_{\theta, 0} = K(\theta)$ and $P_{ \theta, 0} = 0$.  

To solve for $W_{\theta, 1}, P_{\theta, 1}$ we differentiate \eqref{stablemanifold} at a point $(\theta, 0)$ to obtain
\begin{equation}\label{order1}\begin{split}
&DF \circ W_{\theta, 1} = W_{\theta + \omega, 1} P_{\theta, 1} \\
\end{split}\end{equation}
%Recall that the projections $\Pi^s_{K(\theta)}$ satisfy
%\begin{equation}\label{stableinvariance}
%DF(K(\theta))\Pi^s_{K(\theta)} = \Pi^s_{K(\theta + \omega)} DF(K(\theta))
%\end{equation}
To solve \eqref{order1} it suffices to take  $W_{\theta, 1}(s) = (s, 0, 0)$ 
and $P_{\theta, 1} = A^s(\theta )$, both of 
which are in $S^1_{\underline{\rho}_1, \underline{c}, \Gamma}$.  
Note that this choice for $W_{\theta, 1}$ is not unique since, for instance, we could have also chosen 
$W_{\theta, 1}(s) = \sigma (s, 0, 0)$, for non-zero real number $\sigma$.  

For $i > 1$, we will solve for $W_{i, \theta}$, $P_{i \theta}$ inductively.  Taking the $i^{th}$ derivative of \eqref{stablemanifold} we obtain 
\begin{equation}\label{generalderivative}\begin{split}
DF(K(\theta)) W_{\theta, i} + r_i &= W_{\theta + \omega, i} P_{\theta, 1}^{\otimes i} + W_{\theta + \omega, 1} P_{\theta , i} \\
& = W_{\theta + \omega, i} (A^s(\theta))^{\otimes i} + P_{\theta, i}
 \end{split}\end{equation}
where $r_i$ is a polynomial expression in $W_{\theta, j}, P_{\theta, j}$,$j \leq i - 1$, and $F$ and its derivatives up to order $i$.  
The fact that each term in \eqref{generalderivative} belongs to $S^k_{\underline{\rho}_1, \underline{c}, \Gamma}$ is a consequence of Lemma \ref{complineartheta}.
We will consider the projections of the equations onto $\mathcal{E}^{s}_{K(\theta)}$ and $\mathcal{E}^c_{K(\theta)} \oplus \mathcal{E}^u_{K(\theta)}$.  
If we let $W_{i, \theta}^s = \Pi^s_{K(\theta)}W_{i, \theta}, W^{c \oplus u}_{i, \theta} = \Pi^c_{K(\theta)} \oplus \Pi^u_{K(\theta)} W_{i, \theta}$ and 
similarly for $P_i$ and  $r_i$, then the projected equations become

\begin{equation}\label{projected}\begin{split}
&A^s(\theta) W_{i, \theta}^s - W_{i, \theta + \omega}^s (A^s(\theta))^{\otimes i}  - P_{i, \theta } = - r_i^s \\
&A^{c \oplus u}(\theta) W_{i, \theta}^{c \oplus u} - W^{c \oplus u}_{i, \theta + \omega} (A^s(\theta))^{\otimes i} = -r_i^{c \oplus u}
\end{split}\end{equation}

The first of these equations can be solved by taking $W^s_{i, \theta} = 0$ and $P^s_{i, \theta} =  r_i$, while the second equation requires a bit 
more work.  We first start by rewriting the equation as

\begin{equation}\label{rewrite}
\mathcal{L}_{A^{c \oplus u}} W^{c \oplus u}_{i, \theta} - \mathcal{L}^\omega_{i, A^s, \text{Id}} W^{c \oplus u}_{i, \theta} = -r_i^{c \oplus u}
\end{equation}

Thus if we show that $\mathcal{L}_{A^{c \oplus u}} - \mathcal{L}^\omega_{i, A^s, \text{Id}}$ is invertible then we can
conclude that choosing $W^s_{i, \theta} = 0$ allows us to uniquely determine $W^{c \oplus u}_{i, \theta}$.   
Using the general fact that $((\mathcal{L}_B)^{-1} W)(\theta) = (B(\theta))^{-1} W(\theta)$ we have
\begin{equation}\label{almostdone}
\mathcal{L}_{A^{c \oplus u}} - \mathcal{L}^\omega_{i, A^s, \text{Id}} = \mathcal{L}_{A^{c \oplus u}} \left( \text{Id} - \mathcal{L}^\omega_{i, (A^s), (A^{c \oplus u})^{-1}} \right)
\end{equation}
Thus, by Lemma \ref{technicalspectrumlemma} and the assumptions of Theorem \ref{maintheorem_spec} imply that 
$1 \notin \text{Spec}(\mathcal{L}^\omega_{i, (A^s), (A^{c \oplus u})^{-1}})$, it follows that $\mathcal{L}_{A^{c \oplus u}} - \mathcal{L}^\omega_{i, A^s, \text{Id}}$ 
is invertible.
$\Box$

\subsection{Formulation as a Fixed Point Problem}\label{fixed_section1}
In this section we will use the fact that we can write $W = W^\leq + W^>$, where $W^>$ vanishes up to order $L$ in the variable $s$.  
We will solve an equivalent form of \eqref{stablemanifold}, namely if we let $G(\theta, s) = F(K(\theta) + s)$, then we will solve

\begin{equation}\label{stableequation}
G(\theta, \tilde{W}(\theta, s)) = G(\theta , 0) + \tilde{W}(\theta + \omega, P(\theta, s))
\end{equation}
Where $\tilde{W}(\theta, 0) = 0$ for all $\theta$. Note that $\tilde{W}$ solves \eqref{stableequation} if and only if 
$W: = K(\theta) + \tilde{W}$ solve \eqref{stablemanifold}.  The advantage of working with \eqref{stableequation} is that there is a nice scaling of this 
equation that is not present in \eqref{stablemanifold}.  Namely if we consider $G^\delta(\theta, s) := \frac{1}{\delta} G(\theta, \delta s)$ and similarly 
for $\tilde{W}$ and $P$ we have that \eqref{stableequation} holds in 
a $\delta$ neighborhood $B_{\underline{\rho}_1, \delta}$ of the zero section of $\mathcal{E}^s$
 if and only if 
\begin{equation}
G^\delta(\theta, \tilde{W}^\delta(\theta, s)) = G^\delta(\theta, 0) + \tilde{W}^\delta(\theta + \omega, P^\delta(\theta, s))
\end{equation}
holds in unit ball bundle $B_{\underline{\rho}_1, 1}$ of $\mathcal{E}^s$.  
This is motivated by the following observation:
we only want to scale only in the directions not tangent to the torus, that is we only want to scale the $s$ variable. 
This has a natural
 interpretation for $G$, $\tilde{W}$ and $P$ since they formally depends on $s$ and $\theta$ and we can therefore scale only the $s$ variable, but
 since $F$ is formally defined on the entire phase space, any scaling $F^\delta$ of $F$ will results in scaling both $s$ and $\theta$ in the
 composition $F^\delta \circ W^\delta$.  

This choice of scaling also has the following crucial property:  let $N(\theta, s): = G(\theta, s) - G(\theta, 0) - D_s G(\theta, 0) s$.  Then 
$\|  D^i_s N \|_{C^0( B_{\underline{\rho}_1, 1})} \rightarrow 0$ as $\delta \rightarrow 0$
for any $i = 0, 1 \ldots$.  This will allow us to obtain 
better estimates when solving
the fixed point equation we consider.  Note also that the domain of analyticity of $G$ is 
$B_{\underline{\rho}_1, \rho_2}$
while the domain of analyticity was 
$$\{ z \in \mathcal{M} \text{ } | \text { } |z - K(\theta)| \leq \rho_2 \text{ for some }\theta \text{ with }
|\text{ Im } (\theta_n)| < \rho_{1, n} \text{ for all } n \}$$

Note that solving \eqref{stablemanifold} up to order $L$ is equivalent to solving \eqref{stableequation} up to order $L$.  
Thus we can assume that we have $\tilde{W}^{\leq}$ and $P$, polynomials in $s$ that solve \eqref{stableequation} up to order $L$.  Now we will write a fixed
point equation for a function $\tilde{W}^>$ such that $\tilde{W}: = \tilde{W}^{\leq} + \tilde{W}^>$ solves \eqref{stableequation}.  To this end, we note that we have 
the following Taylor expansion for $G$
\begin{equation}\label{TaylorG}
 G(\theta, \tilde{W}(\theta, s)) = G(\theta, 0) + D_sG(\theta, 0)(\tilde{W}^\leq + \tilde{W}^>) + N(\theta, \tilde{W}^\leq + \tilde{W}^>)
\end{equation}
Assuming that $\tilde{W}$ solves \eqref{stableequation} then
\begin{equation}\begin{split}
 &G(\theta, 0) + \tilde{W}(\theta + \omega, P(\theta, s)) \\
&= G(\theta, 0) + D_sG(\theta, 0)(\tilde{W}^\leq + \tilde{W}^>) + N(\theta, \tilde{W}^\leq + \tilde{W}^>)
\end{split}\end{equation}
Rearranging terms and defining $Q_\omega := (\text{Id} , P)\circ (T_\omega, \text{Id})$ we have
\begin{equation}\begin{split}\label{almostfixedpoint}
 & DF(K(\theta))\tilde{W}^> - \tilde{W}^>\circ Q_\omega = \\
 & -N\circ(0, \tilde{W}^\leq + \tilde{W}^>) - DF(K(\theta))\tilde{W}^\leq + \tilde{W}^\leq \circ Q_\omega
\end{split}\end{equation}

If we define the operator $S$ by
\begin{equation}\label{S_operator}
S H := D F(K(\theta))H - H \circ Q_\omega
\end{equation}
then the idea to formulate the problem of finding $\tilde{W}^>$ via a fixed point argument becomes clear: If we can show that $S$ is invertible 
then \eqref{almostfixedpoint} becomes
\begin{equation}\begin{split}\label{fp_eqn}
\tilde{W}^> =  &S^{-1} [- N \circ (\theta, \tilde{W}^\leq + \tilde{W}^>) - D F(K(\theta))\tilde{W}^\leq   + \tilde{W}^\leq \circ Q_\omega]
\end{split}\end{equation}

\subsubsection{The Invertibility of $S$}\label{Section_invS}

We start by defining an appropriate space of functions in which $W^>$ lies to guarantee that $S$ is invertible.  
Given a decay function $\Gamma$ and an positive integer $\ell$ we will consider the norm on functions that vanish up to order $l$ in the variable $s$
\begin{equation}
\|H \|_{\Omega^{\underline{\rho}}_{\underline{c},\Gamma, \ell}} :=
\max_{0 \leq i \leq \ell + 1} \| D_s^{i} H \|_{C_{\underline{c}, \Gamma}^0(B_{\underline{\rho}, 1})}
\end{equation}
and consider the space
\begin{equation}\label{Omega_defn}
\Omega^{\underline{\rho}}_{\underline{c},\Gamma, \ell} = 
\begin{Bmatrix} &H : B_{\underline{\rho}, 1}  \rightarrow \mathcal{M} \text{ } : \text{ } H \text{ is analytic on } B_{\underline{\rho}, 1}, \\ 
&D_s^k H(, \theta, 0) = 0 \text{ for } k\leq 0 \leq \ell \text{ and } \|H \|_{\Omega^{\underline{\rho}}_{\underline{c},\Gamma, \ell}} < \infty
\end{Bmatrix}\end{equation}  
where by $D_s$ we mean the derivative with respect to the $s$ component.    

Before showing that $S$ is invertible we need to state a Proposition stating that the composition of 
function is $\Omega^{\underline{\rho}}_{\underline{c},\Gamma, \ell}$ are in $\Omega^{\underline{\rho}}_{\underline{c},\Gamma, \ell}$. 

Before showing that $S$ is invertible we need to state a Proposition stating that the composition of two functions in $\Omega^{\underline{\rho}}_{\underline{c},\Gamma, \ell}$ is also in $\Omega^{\underline{\rho}}_{\underline{c},\Gamma,\ell}$.

Proposition \ref{compprop} is a special case of the more
general result stated for localized $C^r$ with decay in \cite{FLM1}.  In the 
following proposition the $C^r_{\underline{\rho}_1, \rho_2, \underline{c}, \Gamma}$ is defined
in \cite{FLM1}.
In Section \ref{prelim_section} we defined analytic and $C^1$ function spaces, though did not explicitly define $C^r$ function spaces. In \cite{FLM1}, $C^r$ spaces are considered, and without going into the details, we mention that one has the space $C^r_{\underline{\rho}_1, \rho_2, \underline{c}, \Gamma}$ with a norm given by
$$\| f \|_{C^r_{\underline{\rho}_1, \rho_2,  \underline{c}, \Gamma}} :=  \max_{i = 1, \ldots, r} 
\sup_{(\theta, s) \in B_{\underline{\rho}_1, \rho_2} }  \|D^i f\|_{C^0_{\underline{\rho}_1, \rho_2, \underline{c}, \Gamma}}$$
and refer the reader to Sections $2.7$ - $2.9$ of \cite{FLM1} for details
\begin{prop}\label{compprop}
If $H \in \Omega^{\underline{\rho}}_{\underline{c},\Gamma, \ell}$ and $P$ is an analytic polynomial bundle map,
$ P \in C^r_{\underline{\rho}_1, \rho_2, \underline{c}, \Gamma}$
and $P(B_{\underline{\rho}_1, \rho_2}) \subset B_{\underline{\rho}, 1}$, 
then $H \circ P \in \Omega^{\underline{\rho}}_{\underline{c},\Gamma, \ell}$
and 
\begin{equation}
 \| H \circ P \|_{\Omega^{\underline{\rho}}_{\underline{c},\Gamma, \ell}} \leq 
 C_r(1 + \| P \|_{C^r_{\underline{\rho}_1, \rho_2, \underline{c}, \Gamma}}^r)
 \| H \|_{\Omega^{\underline{\rho}}_{\underline{c},\Gamma, \ell}}
\end{equation}
\end{prop}
$\Box$

As the following lemma shows, the operator $S$
defined in Equation \eqref{S_operator} is invertible on 
$\Omega^{\underline{\rho}}_{\Gamma, \ell}$

\begin{lemma}
Under the assumptions of Theorem \ref{maintheorem}, the operator $S$ preserves the space 
$\Omega^{\underline{\rho}_1}_{\underline{c},\Gamma, L}$ (i.e. $x \in \Omega^{\underline{\rho}_1}_{\underline{c}, \Gamma, L}$ 
implies that $S(x) \in \Omega^{\underline{\rho}_1}_{\underline{c}, \Gamma, L}$).  
Moreover, the map $S: \Omega^{\underline{\rho}_1}_{\underline{c},\Gamma, L} \rightarrow 
 \Omega^{\underline{\rho}_1}_{\underline{c}, \Gamma, L}$ is 
a bounded invertible operator, with a bounded inverse and $\|S^{-1} \|$ can be bounded by a constant independent of the scaling parameter $\delta$.  
\end{lemma}
\textit{Proof:}
We first need to check that $S(H) \in \Omega^{\underline{\rho}_1}_{\underline{c},\Gamma, L}$ if $H \in \Omega^{\underline{\rho}_1}_{\underline{c}, \Gamma, L}$.  It is clear 
that $D_s^i (SH)(\theta, 0) = 0$ for $i = 1, \ldots, L$ since $H$ has this property and $Q(\theta, 0) = 0$.  Moreover,
we claim that the $\Omega^{\underline{\rho}_1}_{\underline{c},\Gamma, L}$ norm of each term
defining $S$, namely $D F(K(\theta))H$ and $H \circ Q_\omega$, are finite.  Indeed,
 $\| D F(K(\theta))H \|_{\Omega^{\underline{\rho}_1}_{\underline{c},\Gamma, L}} < \infty$
 since the multiplicative factor of $DF(K(\theta))$ is independent of $s$ and $\| DF(K(\theta)) \|_{\underline{\rho}_1, \underline{c}, \Gamma} < \infty$.

For the $H \circ Q_\omega$ term  note that  $Q_\omega(\theta, 0) = 0$ and $D Q_\omega (\theta, 0) = A^s(\theta)$, 
which is a contraction.  Hence, for a small enough 
scaling parameter, the image of $Q_\omega$ lies in the domain of $H$. 
and hence Proposition \ref{compprop} implies that 
$H \circ Q_\omega \in \Omega^{\underline{\rho}_1}_{\underline{c},\Gamma, L}$.  

To prove the invertibility of $S$, we solve the equation
\begin{equation}\label{invert_S}
SH = \eta
\end{equation}
where $H$ is the unknown and $\eta$ is known.  Equation \ref{invert_S} is equivalent to

\begin{equation}
H = A(\theta)^{-1}H\circ Q_\omega+ A(\theta)^{-1}\eta
\end{equation}
At least formally, we see that
\begin{equation}\label{Ansatz_S}
H = \sum_{j = 0}^\infty A(\theta)^{-j + 1}\eta \circ Q_\omega^j
\end{equation}
is a solution.  We justify that \eqref{Ansatz_S} is a solution by showing that
\begin{equation}\label{Sestimate}
\sum_{j = 0}^\infty \| A^{-(j + 1)} \eta \circ Q_\omega^j \|_{\Omega^{\underline{\rho}_1}_{\underline{c}, \Gamma, L}} \leq C \| \eta \|_{\Omega^{\underline{\rho}_1}_{\underline{c},\Gamma, L}} 
\end{equation}

By the Faa-di-Bruno formula we have
\begin{equation}\label{faa}\begin{split}
D_s^k(\eta \circ Q_\omega^j) = \sum_{i = 0}^k \sum_{k_1 + \cdots + k_i = k} \sigma^{i, k}_{k_1, \ldots, k_i} ([D_s^i \eta] \circ Q_\omega^j) D_s^{k_1}Q_\omega^j \cdots D_s^{k_i} Q_\omega^j
\end{split}\end{equation}
where $\sigma^{i, k}_{k_1, \ldots, k_i}$ are explicit combinatorial coefficients.  
Using \eqref{faa} with $k = L + 1$ we will obtain the desired estimates.  First, we need to estimate each term and factor appearing in \eqref{faa}.  
Since $Q_\omega(\theta, 0) = 0$ and $DQ_\omega(\theta, 0) = A^s(\theta)$ we can, by our scaling assumptions, choose a scaling parameter
 and $\epsilon > 0$ small enough so that both
\begin{equation}\label{DQest}
\| DQ_\omega \|_{C_{\underline{c}, \Gamma}^0(B_{\underline{\rho}_1, 1})} \leq \|A^s \|_{\underline{\rho}_1, \underline{c}, \Gamma} + \epsilon < 1 
\end{equation}
 and
\begin{equation}\label{DQAest}
 \| DQ_\omega \|_{C_{\underline{c}, \Gamma}^0(B_{\underline{\rho}_1, 1})}^{L + 1} \| A^{-1} \|_{\underline{\rho}_1, \underline{c}, \Gamma}
 \leq (\|A^s\|_{\underline{\rho}_1, \underline{c}, \Gamma} + \epsilon)^{L + 1} \|A^{-1}\|_{\underline{\rho}_1, \underline{c}, \Gamma}
\end{equation}
 Recall that $(\|A^s\|_{\underline{\rho}_1, \underline{c}, \Gamma})^{L + 1} \|A^{-1}\|_{\underline{\rho}_1, \underline{c}, \Gamma} < 1$ by
our assumption on $L$.
Since we want estimates on $D^kQ_\omega^j$ we use the fact that $Q$ is a polynomial in $s$ to say that, for $(\theta, s) 
\in B_{\underline{\rho}_1, \rho_2}$
\begin{equation}\label{DQhigherorderest}
 \| D^k Q_\omega^j(\theta, s) \|_{\underline{\rho}_1, \underline{c}, \Gamma} \leq C_k(\|A^s\|_{\underline{\rho}_1, \underline{c}, \Gamma} + \epsilon)^j, 
\end{equation}
for $ k = 0, 1, \ldots, L 
\text{ and } j = 0, 1, 2, \ldots$
Now we need to estimate the factor $(D^i_s \eta)\circ Q_\omega^j$.  Since $D_s^i \eta(\theta, 0) = 0$ for $i = 0, 1, \ldots, L$ we have, by Taylor's 
Theorem
\begin{equation}\label{Detaest}
 \| D_s^i \eta \|_{C^0_{\underline{\rho}_1, \rho_2, \underline{c}, \Gamma}} 
\leq C \|D^{L + 1}_s \eta\|_{C^0_{\underline{\rho}_1, 1, \underline{c}, \Gamma}} \rho_2^{L - k}
\end{equation}
From \eqref{DQhigherorderest} we deduce that the image of $B_{\underline{\rho}_1, 1}$ under the map $Q_\omega^j$ is contained
in $B_{\underline{\rho}_1, \rho_2}$ where $\rho_2 = (\|A\|_{\underline{\rho}_1, \underline{c}, \Gamma} + \epsilon)^j$. From \eqref{DQhigherorderest} and
\eqref{Detaest} we deduce that
\begin{equation}\label{etaQderiv}
 \| (D^i_s \eta)\circ Q_\omega^j \|_{C^0_{\underline{\rho}_1, 1, \underline{c}, \Gamma}} \leq C\|D^{L + 1}_s \eta\|_{C^0_{\underline{\rho}_1, 1, \underline{c}, \Gamma}} \|
(\|A\|_{\underline{\rho}_1, \underline{c}, \Gamma} + \epsilon)^{(L + 1 - i)j}
\end{equation}
Finally using \eqref{faa}, \eqref{DQhigherorderest} and \eqref{etaQderiv} we have
\begin{equation}\begin{split}
& \| D^{L + 1}( A^{-(j + 1)} \eta Q^j_\omega) \|_{C^0_{\underline{\rho}_1, 1, \underline{c}, \Gamma}} \\
& \leq C \sum_{i = 0}^{L + 1}
 \|D^{L + 1}_s \eta\|_{C^0_{\underline{\rho}_1, 1, \underline{c}, \Gamma}} \| \|A^{-1}\|_{\underline{\rho}_1,\underline{c}, \Gamma}^{j + 1} (\|A\|_{\underline{\rho}_1, \underline{c}, \Gamma} + \epsilon)^{(L + 1 - i)j} (\|A^s\|_{\underline{\rho}_1, \underline{c}, \Gamma} + \epsilon)^{ij} \\
& \leq C \|D^{L + 1}_s \eta\|_{C^0_{\underline{\rho}_1, 1, \underline{c}, \Gamma}}  [\|A^{-1}\|_{\underline{\rho}_1, \underline{c}, \Gamma}(\|A\|_{\underline{\rho}_1, \underline{c}, \Gamma} + \epsilon)^{L + 1}]^{j}
\end{split}\end{equation}
Since $(\|A^s\|_{\underline{\rho}_1, \underline{c}, \Gamma})^{L + 1} \|A^{-1}\|_{\underline{\rho}_1, \underline{c}, \Gamma} < 1$ we conclude that \eqref{Sestimate} holds.

$\Box$

\subsection{Solving the Fixed Point Equation \eqref{fp_eqn}}\label{fixed_section2}
Now that we have shown that $S$ is invertible, recall that we wish to solve the equation
 \begin{equation}\label{fixedpoint}\begin{split}
\tilde{W}^> =  \mathcal{T}(\tilde{W}^>)
\end{split}\end{equation}
where  $\mathcal{T}$ is defined by
\begin{equation}\begin{split}\label{T_def}
\mathcal{T}(\tilde{W}^>) = &S^{-1} [- N \circ (\theta, \tilde{W}^\leq + \tilde{W}^>) - DF(K(\theta))\tilde{W}^\leq +  \tilde{W}^\leq \circ Q_\omega]
\end{split}\end{equation}
The equation \eqref{T_def} is equivalent to the invariance equation \eqref{inv_eqn_thm} for $W = W^\leq + W^>$, see \eqref{fp_eqn}.
We now show that $\mathcal{T}$ is a contraction in $\Omega^{\underline{\rho}_1}_{\underline{c}, \Gamma, L}$ defined in \eqref{Omega_defn}

\begin{lemma}\label{contraction}
Under the assumptions of Theorem \ref{maintheorem} and under the scaling assumptions, $\mathcal{T}$
 sends the closed ball of radius $\frac{\rho_2}{3}$ 
, $\overline{B}_{\frac{\rho_2}{3}}$, of $\Omega^{\underline{\rho}_1}_{\underline{c}, \Gamma, L}$ into itself and is a contraction.  
Therefore $\mathcal{T}$ has a fixed point $W^>$ in 
the closed unit ball of $\Omega^{\underline{\rho}_1}_{\underline{c}, \Gamma, L}$.  
\end{lemma}
\text{Proof:}
First we show that $\mathcal{T}$ maps points in $\overline{B}_{\frac{\rho_2}{3}}$ to 
$\Omega^{\underline{\rho}_1}_{\underline{c}, \Gamma, L}$. The more refined estimate that $\mathcal{T}$ maps
$\overline{B}_{\frac{\rho_2}{3}}$ to itself will be proven later.  
By choosing a small enough scaling parameter, we can
assume that $\tilde{W}^\leq$ is arbitrarily close to the immersion into the stable subspaces $\mathcal{E}^s_{K(\theta)}$ and $Q_\omega$ to be
arbitrarily close to $A^s$.  Thus if $\tilde{W}^>$ is in $\overline{B}_{\frac{\rho_2}{3}}$, 
then the image of $\tilde{W}^\leq + \tilde{W}^>$ lies in the ball of radius $\overline{B}_{\rho_2}$, and recall
that $N(\theta, s)$ is assumed to be analytic on $B_{\underline{\rho}_1, \rho_2}$.  
Thus, by Proposition \ref{compprop} we can conclude that 
$N \circ (\theta, \tilde{W}^\leq + \tilde{W}^>)$ is in $\Omega^{\underline{\rho}_1}_{\underline{c}, \Gamma, L}$.

The other terms $DF(K(\theta))\tilde{W}^\leq$ is in $\Omega^{\underline{\rho}_1}_{\underline{c}, \Gamma, L}$
because the multiplying factor of  $DF(K(\theta))$ does not
depend on $s$, and since $Q_\omega(\theta, 0) = 0$ and
$DQ_\omega(\theta, 0) = A^s(\theta)$ it follows that  $W^\leq \circ Q_\omega$
is also in  $\Omega^{\underline{\rho}_1}_{\underline{c}, \Gamma, L}$.

We show that $\mathcal{T}$ is a contraction.  For $\tilde{W}^>$ and $\tilde{W}^> + \Delta$ in the closed until ball of $\Omega^{\underline{\rho}_1}_{\underline{c}, \Gamma, L}$.
We have the increment formula
\begin{equation}\label{increment}\begin{split}
\mathcal{T}(\tilde{W}^> + \Delta) - \mathcal{T}(\tilde{W}^>) &= \int_0^1 \frac{d}{d\tau}[\mathcal{T}(\tilde{W}^> + \tau \Delta)] d\tau\\
& = -\int_0^1 S^{-1} D_sN (\theta, \tilde{W}^\leq + \tilde{W}^> + \tau \Delta)\Delta d\tau
\end{split}\end{equation} 
Taking the $(L + 1)$-derivative of \eqref{increment} in the $s$ variable we obtain
\begin{equation}\label{contr}
\| \mathcal{T}(\tilde{W}^> + \Delta) - \mathcal{T}(\tilde{W}^>) \|_{\Omega_{\underline{c}, \Gamma, L}^{\underline{\rho}_1}} \leq C \| N \|_{C_s^{L + 2}}
 \| \Delta \|_{\Omega_{\underline{c}, \Gamma, L}^{\underline{\rho}_1}}
\end{equation}
Since $\| N \|_{C_s^{L + 2}} \rightarrow 0$ as the scaling parameter goes to zero, we conclude that $\mathcal{T}$ is a contraction for sufficiently
small values of the scaling parameter.

Now we show that $\mathcal{T}$ maps $\overline{B}_{\frac{\rho_2}{3}} \subset \Omega^{\underline{\rho}_1}_{\underline{c}, \Gamma, L}$ into itself.  
For $\| \tilde{W}^> \|_{\Omega^{\underline{\rho}_1}_{\underline{c},\Gamma, L}} \leq 1$ we have
\begin{equation}\begin{split}
& \mathcal{T}(\tilde{W}^>) = \mathcal{T}(0) + ( \mathcal{T}(\tilde{W}^>) - \mathcal{T}(0)) \\
&= (S^{-1}[-G(\theta, \tilde{W}^\leq) + G(\theta, 0) + \tilde{W}^\leq \circ Q_\omega] ) +(\mathcal{T}(\tilde{W}^>) - \mathcal{T}(0))
\end{split}\end{equation}
Since $\| S^{-1} \|$ is independent of the scaling parameter, we can say that the first term can be made as 
small as we wish with a small
enough scaling parameter,
 and the second term has norm smaller than $\frac{\rho_2}{3}$ since $\tilde{W}^>$ does and $\mathcal{T}$ 
is a contraction.  

$\Box$

\subsection{Proof of Lemma \ref{uniqueness_lem}}\label{uniqueness_section}
%In this section we prove the uniqueness results stated in Section \ref{uniqueness_section}.  The two results 
%are Theorem \ref{uniqueness_thm} that gives conditions for $W$ and $P$ to guarantee uniqueness and 
%Proposition \ref{w_reparam} that relates a given $\tilde{W}$ to the $W$ of Theorem \ref{uniqueness_thm}
%by a reparameterization of the $s$ variable.  

%\subsubsection{Proof of Theorem \ref{uniqueness_thm}}
In this section we prove the Lemma \ref{uniqueness_lem} stated in Section \ref{uniqueness_statement_section}.  In the proof of Proposition \ref{induction}, we saw that there was a lack of uniqueness in solving for $W_{i, \theta}$.
However once we chose $W_{i, \theta}$ and $P_{i, \theta}$ for $1 = 1, \ldots, L$, then using
a fixed point argument we constructed $W^>$ such that the pair $(W, P)$ where $W$ is given by
\begin{equation}
 W = \sum_{i = 1}^L W_{i, \theta} + W^>
\end{equation}
 satisfies the conclusions of Theorem \ref{maintheorem_spec}.  Since $W^>$ was constructed using fixed point 
 argument, it is unique once the low order terms of $W^\leq$ and $P$ have been specified.  
 That conditions $1$-$4$ of Lemma 
\ref{uniqueness_lem} guarantee that $W$ and $P$ thus follows from the proof of Proposition \ref{induction}
and the fact that $W^>$ is unique.  

Now we want show the uniqueness of the manifold itself.
Let $W$ be the analytic solution of Equation \eqref{stablemanifold} constructed from Theorem \ref{maintheorem_spec}
 and write
\begin{equation}
 W_{\theta} = (W^s_{\theta}, W^{c \oplus u}_{\theta}) = (\Pi_{K(\theta)}^s W_{\theta}, \Pi_{K(\theta)}^{c \oplus u}W_{\theta})
\end{equation}
Moreover, consider the Taylor series up to order $L$ and write  $W_{\theta} = \sum_i^L W_{i, \theta} + W^>$.  
If we suppose that $W^s_{1, \theta} = \Pi_\theta^s$ and $W^s_{i, \theta} = 0$ for $i = 2, \ldots, L$, then the proof
of Proposition \ref{induction} implies that this choice of $W^s_{i, \theta}, i = 1, \ldots, L$ determines uniquely 
$W^{c \oplus u}_{i,\theta}$ for $i = 1, \ldots, L$, and the proof of Lemma \ref{fixedpoint} then implies that $W^>$ is
also determined uniquely by $W^s_{i, \theta}, i = 1, \ldots, L$. 

We use this observation to prove that the manifold $W(\{\theta \} \times B^s_1)$ constructed in the proof of Theorem \ref{maintheorem} is unique among all manifolds that are invariant under $F$, tangent to $\mathcal{E}^s_{K(\theta)}$ and admit an analytic parameterization that decay like $\Gamma$.  
Indeed, suppose that $V_\theta \in \mathcal{A}_{\underline{\rho}_1, \rho_2, \underline{c}, \Gamma}$ is an embedding of an invariant
manifold, the image of $D V_\theta (0)$ is $\mathcal{E}^s_{K(\theta)}$ and the dependence on $\theta$ is analytic.  If we write 
$V_\theta = (V^s_\theta, V_\theta^{c\oplus u})$then by the implicit function theorem we know that $V^s_\theta$ is invertible in a neighborhood of 
the origin.  Thus if we let $H = V_\theta^{c\oplus u} \circ (V^s_\theta)^{-1}$, then the image of $V_\theta$ is the same as the graph of $H$. 
 Since the graph of $H$ is invariant we can conclude that
\begin{equation}
 F^{c \oplus u} \circ (\text{Id}, H) = H \circ F^s \circ (\text{Id}, H)
\end{equation}
Thus, it follows that Equation \eqref{stablemanifold} holds if we take $W = (\text{Id}, H)$ and $R = F^s \circ W$.  Moreover, $W = (\text{Id}, H)$ satisfies $W^s_{1, \theta} = \Pi_\theta^s$ and $W^s_{i, \theta} = 0$ for $i = 2, \ldots, L$ and hence $W$ coincides with the solution given in the
proof of Theorem \ref{maintheorem}.

\section*{Acknowledgements}
We thank Prof. A. Blass for comments relevant the Lemma \ref{technicalspectrumlemma}.  We also
thank Prof. T. Blass for several conversations and encouragement. We are grateful for Prof. D. Knopf's dedicated work as graduate advisor, which enabled us to collaborate. 
D. Blazevski acknowledges the hospitality of the School of Mathematics at the
Georgia Institute of Technology during the Spring 2012 semester.  
Both authors have been supported by NSF grant DMS1162544 and the Texas Coordinating Board ARP0223
\bibliographystyle{alpha} 
\bibliography{whiskered}

\end{document}